\documentclass[11pt]{article}
\usepackage{mathptmx}
\usepackage[T1]{fontenc}
\usepackage{pifont}
\usepackage{amsfonts}
\usepackage{amsmath, mathptmx}
\usepackage{amssymb}
\usepackage{amsthm}
\usepackage{mathrsfs}
\usepackage{graphics}
\usepackage{lscape}
\usepackage{changebar}
\usepackage{graphicx}
\usepackage{epsfig}
\usepackage{bm}
\usepackage{color}

\makeatletter

\newcommand{\tr}{\hbox{tr}}

\makeatletter

\newcommand{\B}[1]{{\bm{#1}}}

\renewcommand{\=}{\!=\!}

\begin{document}

\title{The fracture of highly deformable soft materials:\\ A tale of two length scales}
\author{Rong Long,$^1$ Chung-Yuen Hui,$^{2,3}$ Jian Ping Gong,$^{3,4,5}$ and Eran Bouchbinder$^{6,*}$\\\\
\small{$^1$Department of Mechanical Engineering, University of Colorado Boulder,}\\
\small{Boulder, CO 80309, USA}\\
\small{$^2$Department of Mechanical and Aerospace Engineering, Field of Theoretical}\\
\small{and Applied Mechanics, Cornell University, Ithaca, NY 14853, USA}\\
\small{$^3$Soft Matter GI-CoRE, Hokkaido University, N21W11, Kita-ku,}\\
\small{Sapporo 001-0021, Japan}\\
\small{$^4$Faculty of Advanced Life Science, Hokkaido University, N21W11,}\\
\small{Kita-ku, Sapporo 001-0021, Japan}\\
\small{$^5$Institute for Chemical Reaction Design and Discovery (WPI-ICReDD),}\\
\small{Hokkaido University, N21W11, Kita-ku, Sapporo 001-0021, Japan}\\
\small{$^6$Chemical and Biological Physics Department, Weizmann Institute of Science,}\\
\small{Rehovot 7610001, Israel}\\\\
\small{$^*$Email: Eran.Bouchbinder@weizmann.ac.il}}

\date{}

\maketitle

\begin{abstract}
The fracture of highly deformable soft materials is of great practical importance in a wide range of technological applications, emerging in fields such as soft robotics, stretchable electronics and tissue engineering. From a basic physics perspective, the failure of these materials poses fundamental challenges due to the strongly nonlinear and dissipative deformation involved. In this review, we discuss the physics of cracks in soft materials and highlight two length scales that characterize the strongly nonlinear elastic and dissipation zones near crack tips in such materials. We discuss physical processes, theoretical concepts and mathematical results that elucidate the nature of the two length scales, and show that the two length scales can classify a wide range of materials. The emerging multi-scale physical picture outlines the theoretical ingredients required for the development of predictive theories of the fracture soft materials. We conclude by listing open challenges and future investigation directions.
\end{abstract}

\section{Introduction: Highly deformable soft materials and fracture as a multi-scale problem}
\label{sec:intro}

Soft elastomers or gels, featuring a low shear modulus (in the kPa-MPa range, in contrast to conventional stiff materials with moduli in the GPa range), are emerging as functional components in many engineering applications. Their capability to undergo large, reversible deformation offers unique opportunities for technological fields such as robotics and electronics, e.g.~as manifested in fast-growing research on soft robotics~\cite{Polygerinos2017,Cianchetti2018,Wallin2018} and stretchable electronics~\cite{Rogers2010,Lin2016,Yang2018}. Moreover, soft materials can be engineered to be compatible with biological cells or tissues, both chemically and mechanically, and thus are inherently advantageous in biomedical engineering applications. Examples include, among others, tissue engineering scaffold~\cite{Drury2003}, artificial cartilage~\cite{Yasuda2005}, contact lens~\cite{Stapleton2006} and biomedical adhesives~\cite{Li2017, Blacklow2019}. These rapid technological developments are accompanied by pressing scientific questions about the basic physics of these materials. Most notably, a fundamental question that arises concerns the ways in which such soft materials sustain large deformation without failure. Deeply understanding the underlying physics can not only facilitate the development of theoretical approaches to predict failure in soft materials and to better characterize them using careful laboratory experiments, but also establish physical guiding principles to enhance their mechanical robustness.

The mechanical failure of solids is a long-standing problem in physics~\cite{Orowan1949, freund1998dynamic, Broberg.99,BFM10,Bouchbinder2014}, which is an intrinsically complex phenomenon that couples physical processes at length and time scales that are separated by many orders of magnitude, giving rise to a wealth of emergent behaviors. Theoretically speaking, the ideal strength of a material, i.e.~the maximum stress that it can withstand, might be estimated based on the energy invested in breaking individual atomistic bonds. Consequently, following Orowan~\cite{Orowan1949}, the ideal/theoretical strength $\sigma_m$ of brittle solids can be estimated by comparing the linear elastic strain energy density $\sigma_m^2/E$, where $E$ is Young's (extensional) modulus, to the ratio of the bare surface energy $\gamma$ and the atomistic separation length $a_0$, $\gamma/a_0$. The resulting ideal/theoretical strength estimate, $\sigma_m\!\sim\!\sqrt{\gamma E/a_0}$, assumes that the material deformation is predominantly linear elastic and homogeneous such that the macroscopically applied stress $\sigma_m$ is transmitted to the atomistic scale. The ideal/theoretical strength, however, is rarely achieved in practice. For example, the ideal/theoretical strength of window glass is estimated to be on the order of $10$ GPa based on Orowan's relation, which is several orders of magnitude higher than the actual (measured) strength of glass, which is found to be in the range $0.01\!-\!0.1$ GPa~\cite{Littleton1923, Kasunic2015}. This huge discrepancy is attributed to the almost inevitable existence of defects (e.g., voids or micro-cracks) at scales larger than the atomistic scale $a_0$, as first pointed out by Griffith~\cite{Griffith1921}. To illustrate the dramatic effect of defects on the strength of materials, Griffith measured the tensile strength of glass fibers of different diameters~\cite{Griffith1921} and found that the strength drastically increases from $\sim\!0.2$ GPa to $\sim\!3$ GPa when fiber diameter was reduced from $\sim\!1$ mm to $\sim\!3\,\mu$m, which he attributed to the reduction in defect number and size in the thinner fibers. It is now well-established that defects, in particular their geometry and typical size $c$, give rise to stress concentration and hence to the initiation of localized material failure in their vicinity. That is, it is well-accepted that the homogeneous stress assumption generically breaks down and that macroscopic stresses are strongly amplified by material defects. Indeed, a more realistic estimate of the strength $\sigma_m$ is obtained once the atomistic length $a_0$ in Orowan's relation is replaced by the defect length $c$, resulting in Griffith's relation $\sigma_m\!\sim\!\sqrt{\gamma E/c}$~\cite{Griffith1921}, which can be significantly smaller than the ideal/theoretical strength, if $c\!\gg\!a_0$.

\begin{figure}[ht!]
\begin{centering}
\includegraphics[width=\textwidth]{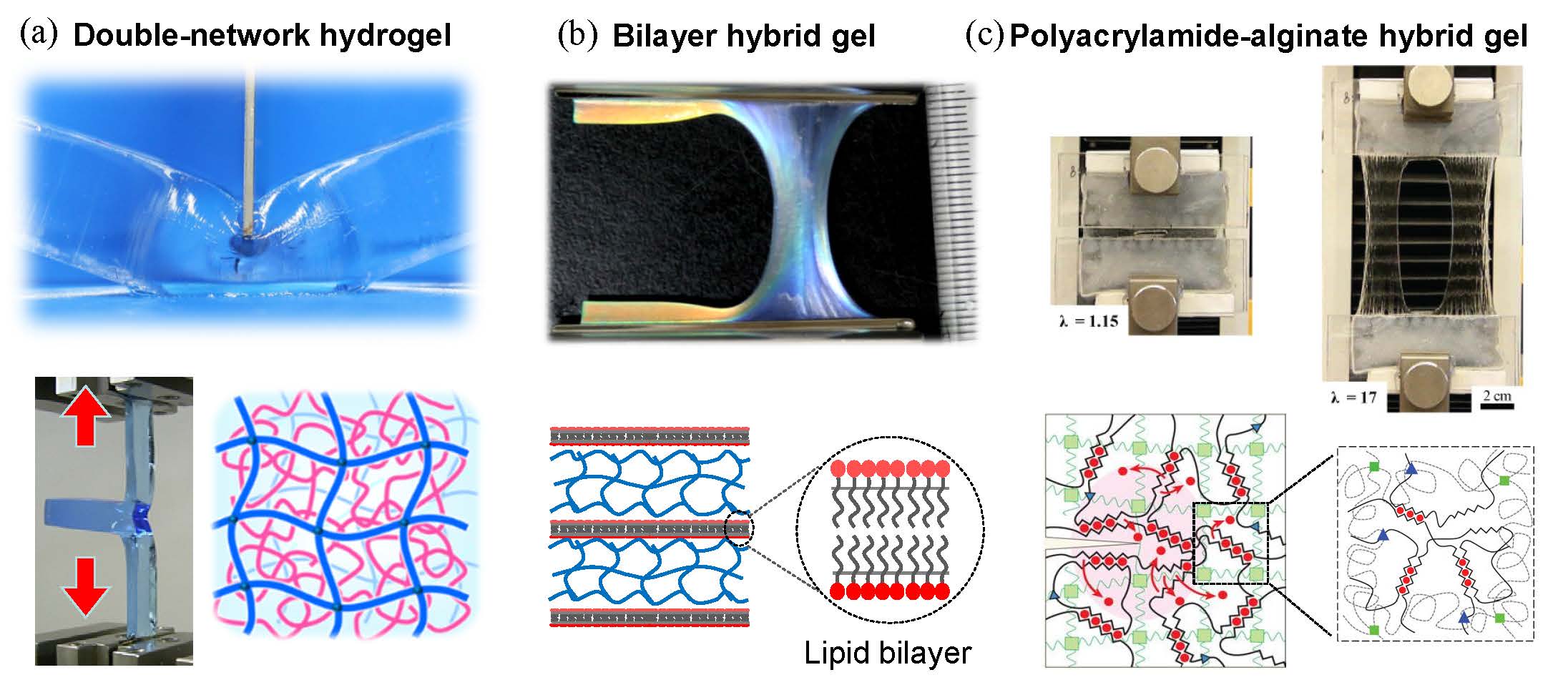}
\par\end{centering}
\caption{Experimental examples of the failure resistance of tough, highly deformable soft materials. For each material, an illustrative sketch of the underlying molecular architecture is added (the reader is referred to the accompanying reference for more details). (a) The cutting (upper panel) and tearing (lower panel) resistance of double-network (DN) gels~\cite{Gong2003,Gong2010}. (b) The severe blunting of an originally sharp crack in a bilayer hybrid gel~\cite{Haque2010, Haque2011} under very large stretch is demonstrated. (c) The failure resistance of a notched polyacrylamide-alginate hybrid gel is demonstrated~\cite{Sun2012}. The notched sample stretched by $15\%$ of its initial length (a stretch of $\lambda\!=\!1.15$, left) and stretched to $17$ times its initial length (a stretch of $\lambda\!=\!17$, right).\label{fig:exp}}
\end{figure}

The initiation of localized material failure in the vicinity of defects --- which involves various new length scales --- leads in many cases to the formation of rather sharp cracks that extend and propagate into the solid. When cracks propagate throughout a solid (of typical linear size $L$), catastrophic failure is induced and the solid completely loses its macroscopic load bearing capacity. The resistance to crack initiation and propagation has been recognized as a critical material property, involving complex physics at various length scales, and has become the central topic of fracture mechanics~\cite{Anderson2017, Zehnder2012}. Studies on the fracture of soft materials date back to 1950's, when Rivlin and Thomas~\cite{Rivlin1953} pioneered the research on the failure of rubber due to its industrial importance. Interest in this field has been renewed, and in fact significantly expanded, in recent years due to the growing range of technological applications of soft materials. Soft materials --- such as rubber, elastomers and gels --- can in fact exhibit vastly different fracture behaviors depending on their small-scale physics, i.e.~microstructure or molecular architecture. For example, hydrogels consisting of a crosslinked polymer network swollen by water are typically brittle~\cite{Zhao2014,Creton2016}, as reflected in the sensitivity to defects and unstable crack propagation observed in such gels. In contrast, double-network (DN) gels, consisting of a stiff swollen network interpenetrating with a soft extensible network, exhibit substantially enhanced resistance to crack propagation~\cite{Gong2003, Gong2010}. Some of the remarkable failure resistance properties of highly deformable soft materials are illustrated in {\bf Figure~\ref{fig:exp}}.

Yet, whether brittle or tough, soft materials share a common feature that distinguishes them from ordinary stiff materials such as glass, ceramics and metals: they all feature strongly nonlinear strain and stress fields near crack tips. Conventional fracture mechanics theory, developed mainly for stiff materials, is based on infinitesimal strains and has been deemed inadequate for soft material fracture~\cite{Livne2010, Lefranc2014}. In particular, the nonlinear crack tip fields in soft materials can be coupled to small-scale molecular failure processes~\cite{Lake1967, Baumberger2006a, Baumberger2006}, mesoscale energy dissipation~\cite{Brown2007, Tanaka2007, Zhang2015}, and larger scale effects such as crack blunting~\cite{Hui2003,Seitz2009}. A thorough understanding of the fracture behavior of soft materials is thus an intrinsically multi-scale physical problem, which requires both accurate quantitative experimental characterization and theoretical analysis of nonlinear crack-tip mechanics.

The surge of interest in soft material fracture has also been reflected in several recent review articles~\cite{Zhao2014,Creton2016,Long2015,Long2016,Bai2019}. In~\cite{Creton2016} a rather comprehensive overview of the unique fracture behaviors of soft materials has been presented. Designing tough hydrogels by controlling the dissipation associated with their underlying networks has been reviewed in~\cite{Zhao2014}, and measurements and quantitative interpretation of the fracture toughness of these hydrogels have been reviewed in~\cite{Long2016}. Large strain crack tip effects, as manifested in the nonlinear, quasi-static asymptotic solutions have discussed in~\cite{Long2015}. Finally, experimental progress in relation to fatigue crack propagation in soft hydrogels under cyclic loading has been recently summarized in~\cite{Bai2019}. The goal of the present review is to complement these by offering a physics-oriented perspective on the multi-scale nature of the fracture of soft materials, aimed mainly --- but not exclusively --- at physicists. In particular, we focus on the important roles played by two length scales that highlight soft material fracture: a length scale associated with large elastic deformation near crack tips and a length scale associated with near tip dissipation.

As will be elaborated on below, these two length scales allow us to classify a wide range of materials, featuring a broad range of deformation and failure behaviors. More importantly, these length scales provide a unified picture of crack tip physics, thereby outlining the ingredients required for the development of predictive theories of the fracture of soft materials.

\section{Conventional linear elastic fracture mechanics}
\label{sec:LEFM}

To set the stage for the discussion of the fracture of highly deformable soft materials to follow, we briefly review here the main elements of the conventional theory of fracture, which has been mainly designed for stiff brittle materials~\cite{Anderson2017, Zehnder2012}. The theory, termed Linear Elastic Fracture Mechanics (LFFM), is based on two major assumptions. First, as the name implies, this theory assumes that the material behavior is predominantly linear elastic prior to failure. That is, it is assumed that the deformation of a body containing a crack --- quantified by the displacement vector field ${\B u}$ --- gives rise to stresses (forces per unit area) --- quantified by the Cauchy stress tensor field $\B \sigma$ --- that are linearly related to the gradient of the displacement, ${\B \nabla}{\B u}$ (various strain measures can be defined using the displacement gradient). As such, LEFM is a perturbative approach that is based on the leading order expansion in the smallness of ${\B \nabla}{\B u}$. Second, it is assumed that linear elasticity breaks down in a negligibly small region near the crack tip, where nonlinearity, dissipation and material failure take place (the so-called fracture process zone). The dissipation in this region is quantified by the fracture energy $\Gamma$, which represents the energy dissipated during crack advance per unit area.

In terms of Orowan's relation in the ideally brittle limit, discussed above in Section~\ref{sec:intro}, the fracture energy $\Gamma$ is identified with the bare surface energy $\gamma$ and dissipation takes place over an atomistic length scale $a_0$. These assumptions may remain valid when the fracture energy $\Gamma$ is in fact larger than the surface energy $\gamma$ (e.g.~due to plastic deformation), which also implies that the dissipation occurs on a length scale larger than $a_0$ (e.g.~when a plastic zone develops), as long as the dissipation length is significantly smaller than all other lengths in the problem, most notably the crack size $c$ and the system size $L$. This assumption is sometimes termed the ``small-scale yielding'' condition, where ``yielding'' refers to the onset of plastic deformation and $a_0$ is redefined as the size of the plastic zone. Under these conditions, LEFM makes no reference to the fracture process zone, which is assumed to be a point-like region, and the fracture energy $\Gamma$ is assumed to be an additional input (obtained from experiments or from other theories). As such, LEFM is a scale-free theory that can feature only extrinsic/geometric length scales such as $c$ and $L$.
\begin{figure}
\begin{centering}
\includegraphics[width=0.7\textwidth]{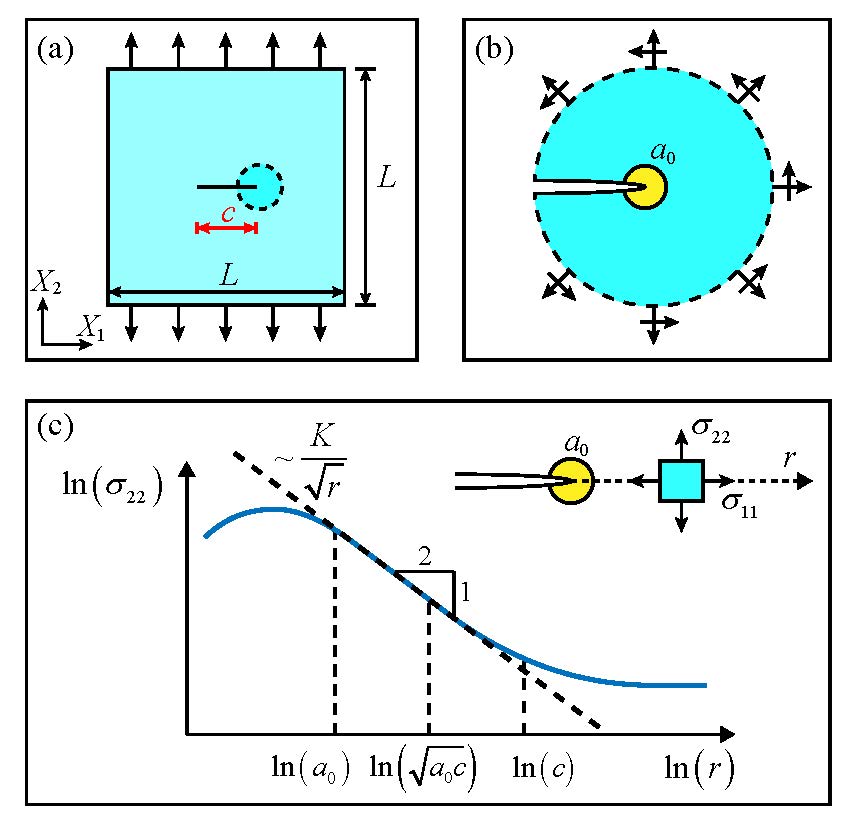}
\par\end{centering}
\caption{Linear elastic fracture mechanics (LEFM) and the asymptotic singular $K$-field. (a) A fracture specimen of linear size $L$ containing a central crack of length $c\!\ll\!L$. The specimen is loaded by uniaxial tension, denoted by the outgoing arrows at the lower and upper boundaries. A region near one of the crack tips, which is zoomed-in on in panel (b), is encircled by a dashed line. (b) A zoom-in on the crack tip region (see panel (a)), showing the stresses that are transferred from the far-field loading to the tip region (represented by the arrows) and a failure zone of linear size $a_0$ (see text for details). (c) A schematic representation of the LEFM tensile (opening) stress $\sigma_{22}$ versus the distance $r$ along the crack line is shown in a log-log scale. For distances larger than the crack length $c$, $r\!\gg\!c$, $\sigma_{22}$ is controlled by the far-field loading. As the tip is approached, $\sigma_{22}$ is amplified due to the presence of the crack. At a distance $\sim\!\sqrt{a_0 c}$ the stress is dominated by the asymptotic singular $K$-field, $\sigma_{22}\!\sim\!K/\sqrt{r}$ (dashed line). At distances smaller than $a_0$, $r\!\ll\!a_0$, the singularity is regularized. (inset, top right) The asymptotic singular $K$-field in quasi-static LEFM predicts equi-biaxial stress conditions ahead of the crack tip, $\sigma_{22}\!=\!\sigma_{11}$. \label{fig:LEFM}}
\end{figure}

With these two assumptions, LEFM makes a few powerful predictions~\cite{Anderson2017, Zehnder2012}. Most notably, LEFM predicts that the stress field around crack tips follows a universal singularity of the form ${\B \sigma}\!\sim\!K/\sqrt{r}$, where $r$ is the distance from the tip and $K$ is the intensity of the singularity, known as the stress intensity factor (a tensorial function of the azimuthal angle is omitted here). From a formal/mathematical perspective, this universal $K$-field is an intermediate asymptotics valid for $a_0\!\ll\!r\!\ll\!c,L$, where information regarding the large scales properties of the problem (e.g.~the loading conditions and geometrical configuration) is transmitted to the fracture process zone through the stress intensity factor $K$. Indeed, for $c\!\ll\!L$ (i.e.~a finite crack in a large body under homogeneous stress far-field loading, cf.~{\bf Figure~\ref{fig:LEFM}a}) one obtains $K\!\sim\!\sqrt{c}$ and for $c\!\gg\!L$ (i.e.~a long crack in a strip of height $L$ under homogeneous displacement loading at the strip edges) one obtains $K\!\sim\!\sqrt{L}$. Moreover, the spatial range of validity of the $K$-field, the so-called $K$-dominant region, can be estimated based on the distance to the crack tip at which the $K$-field is most accurate~\cite{Hui1985}, i.e.~$\sim\sqrt{a_0 c}$ for $c\!\ll\!L$ and $\sim\sqrt{a_0 L}$ for $c\!\gg\!L$. These concepts are illustrated in {\bf Figure~\ref{fig:LEFM}}.

The universal singular $K$-field of LEFM is associated with a finite flux $G$ of elastic energy per unit cracked area, $G\!\sim\!K^2/E$ ($E$ is Young's modulus introduced above). Consequently, crack initiation corresponds to an energy balance of the form $G\!\sim\!K^2/E\!\sim\!\Gamma$, in which elastic energy stored on large scales flows into the fracture process zone, where it is dissipated on small scales as quantified by the fracture energy $\Gamma$. This fundamental LEFM relation further highlights the basic role played by $K$ in coupling the vastly different scales emerging in a fracture problem. Moreover, it shows that the fracture energy $\Gamma$ in fact serves as a threshold for crack initiation, i.e.~it is a material property that quantifies the resistance to crack initiation, since $K$ needs to become sufficiently large such that $G$ first reaches $\Gamma$. Let us apply the energy balance relation $G\!\sim\!K^2/E\!\sim\!\Gamma$ to a crack of length $c$ in a large body under a remote tensile stress $\sigma$, assuming also $\Gamma\!\sim\!\gamma$. For this configuration we have $K\!\sim\!\sigma\sqrt{c}$, which upon substitution in the energy balance relation yields the crack initiation threshold $\sigma_m\!\sim\!\sqrt{\gamma/E c}$. This is nothing but the Griffith prediction discussed in Section~\ref{sec:intro}.

The universal $K$-field also determines the opening profile of a crack near its tip, i.e.~the so-called crack tip opening displacement (CTOD); that is, while cracks in LEFM are assumed to feature no finite radius of curvature in the undeformed/relaxed state (i.e.~assumed to be sharp, not blunted), when opening (tensile) stresses are applied, cracks open up parabolically, with a curvature determined by $K$. Next, we will see that when highly deformable soft materials are considered, the scale-free LEFM framework --- featuring only extrinsic/geometric length scales such as $c$ and $L$ --- breaks down due to the intervention/emergence of new length scales not discussed so far.

\section{Two basic length scales in the fracture of highly deformable soft materials}
\label{sec:2lengths}

The powerful and elegant predictions of LEFM, briefly reviewed above, are based on linear elasticity, i.e.~on a perturbative approach restricted to small reversible deformation. Yet, to understand the fracture of highly deformable soft materials, one needs to consider large (nonlinear) deformation and irreversibility (dissipation), which significantly complicate and enrich the physical picture. In particular, elastic nonlinearity and dissipation are associated with two distinct physical length scales that play important roles in the fracture of highly deformable soft materials.

Highly deformable soft materials, as the name implies, feature large deformation prior to failure. Consequently, the elastic fields surrounding crack tips in such materials can significantly deviate from the universal $K$-fields of LEFM over extended regions. These strongly nonlinear fields, which entail a non-perturbative approach that is highly involved from the mathematical perspective, will be discussed in Section~\ref{sec:asymptotic} below. Here we first consider the following question: at what length scale away from a crack tip the deformation becomes significantly nonlinear elastic? One way to approach this question is to consider situations in which the $K$-field still exists, but crosses over to a nonlinear elastic behavior at smaller $r$'s. This situation is addressed by the weakly nonlinear theory of fracture~\cite{Bouchbinder2014,Livne2010,Bouchbinder2009, Livne.08,Bouchbinder.08}, which is a perturbative approach that takes into account the first nonlinear correction to LEFM. In LEFM, as discussed in Section~\ref{sec:LEFM}, the singular part of the displacement gradient takes the form  ${\B \nabla}{\B u}\!\sim\!(K/E)/\sqrt{r}$. The weakly nonlinear theory predicts that the leading nonlinear correction to this result scales as $(K/E)^2/r$ (the prefactor, which is not discussed here, involves higher order elastic constants). The two contributions become comparable at $r\!\sim\!K^2/E^2$, which together with $K^2/E\!\sim\!\Gamma$ upon crack initiation, imply that the crossover to a nonlinear elastic behavior occurs at a length scale $\sim\!\Gamma/E$.

While the estimate just presented is based on a perturbative approach, we adopt its outcome in a more general context, i.e.~including in situations in which the $K$-field has no range of validity at all, and define the nonlinear elastic length scale $\ell$ as
\begin{equation}
\label{eq:ell}
\ell \sim \Gamma/E \ .
\end{equation}
The length in Equation~\ref{eq:ell} was termed the ``elasto-adhesive'' length in~\cite{Creton2016} and was used to characterize morphology of soft adhesives during debonding~\cite{Shull2004}. Physically, as explained above, $\ell$ represents the typical distance from a crack tip below which the deformation is dominated by elastic nonlinearity at the onset of crack initiation. As the crack tip is further approached, there exists a sufficiently small $r$ at which dissipation sets in. How can one estimate this dissipation length scale?

\begin{figure}
\begin{centering}
\includegraphics[width=\textwidth]{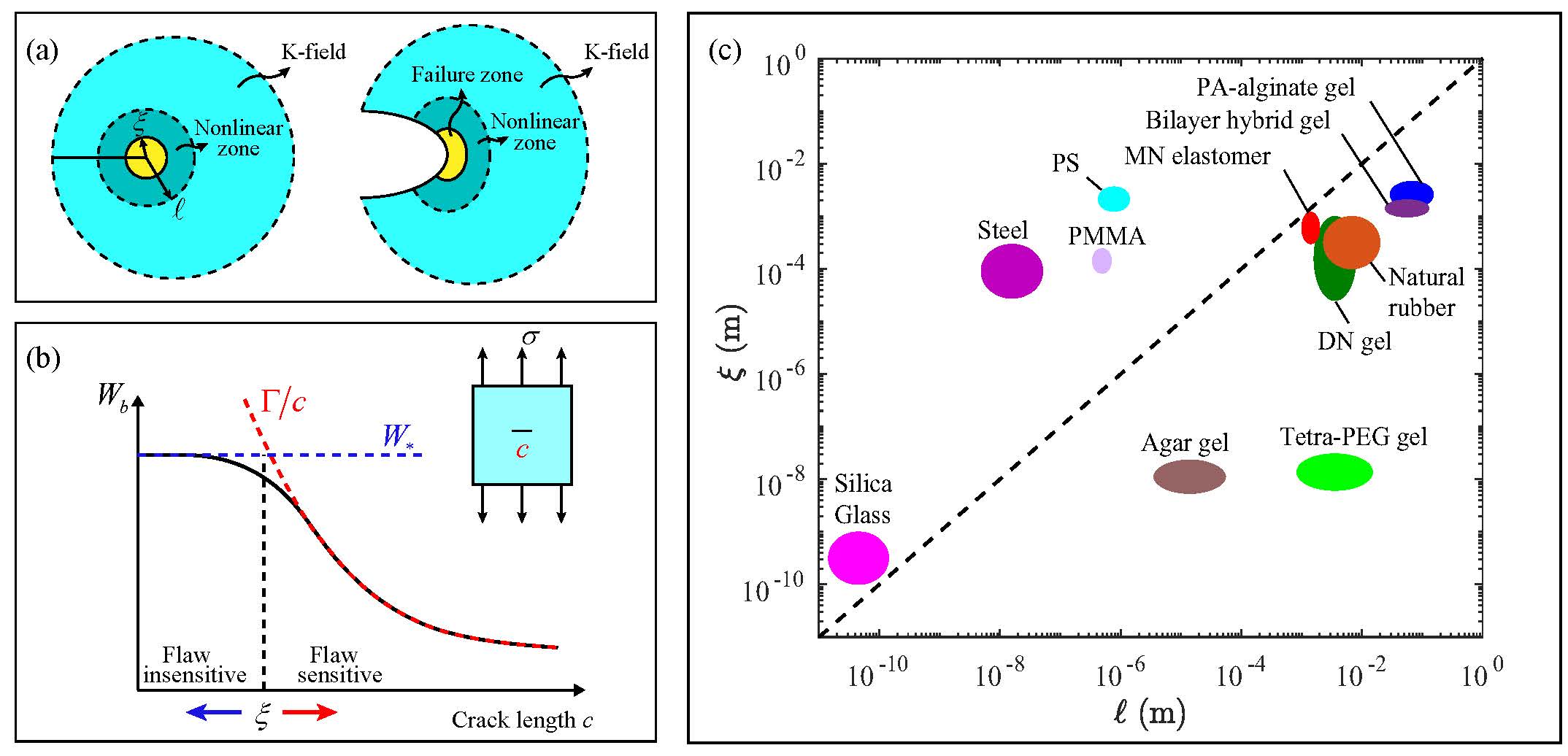}
\par\end{centering}
\caption{Two intrinsic, fracture-related length scales. (a) The nonlinear length scale $\ell$ describes the size of a near crack tip zone where nonlinear elastic effects, due to large deformation, are dominant. The dissipative length scale $\xi$ describes the crack tip failure zone (a generalization of $a_0$ in Figure~\ref{fig:LEFM}). Both $\ell$ and $\xi$ are defined in the reference (undeformed) configuration (left part, the deformed configuration is illustrated on the right part). (b) An illustration of the concept of flaw-insensitivity, achieved by plotting the work per unit volume required to break a large solid $W_b$, containing a central crack with length $c$, versus $c$ (solid line). For $c\!\gg\!\xi$, dimensional analysis implies $W_b\!\sim\!\Gamma/c$ (dashed red line, see the definition of $\xi$ in Equation~\ref{eq:W}), i.e.~$W_b$ is inversely proportional to $c$ (the larger the crack/flaw, the easier it is to break the solid). $W_b$ is bounded from above by $W_*$, which is the work per unit volume required to break a solid in the absence of macroscopic cracks, $c\!\ll\!\xi$ (horizontal dashed blue line). Consequently, $\xi$ marks the crossover from flaw-insensitive failure ($c\!\ll\!\xi$) to flaw-sensitive failure ($c\!\gg\!\xi$), see text for additional discussion. (c) Rough quantitative estimates for $\xi$ and $\ell$ for various materials, with a focus on highly
deformable soft materials. Each material (the names are indicated on the plot), is represented by a colored elliptical blob in the $\xi\!-\!\ell$ plane, where principal axes of each ellipse roughly represent the uncertainty in the available numbers (the uncertainty might in fact be even larger due to the scaling nature of the definitions of $\xi$ and $\ell$, see text for additional discussion). Note also that for soft materials we have $\xi\!<\!\ell$ (in fact, in many cases $\xi\!\ll\!\ell$).\label{fig:2length}}
\end{figure}

To address this question, one usually invokes some typical physical quantity that characterizes dissipative deformation. To keep things as general as possible (i.e.~not to focus on a specific dissipation mechanism), we follow the quasi-brittle fracture theory of Ba\v{z}ant~\cite{Bazant1997}, where it is postulated that there exists a critical energy per unit volume for material failure, $W_*$. Using $\Gamma$ and $W_*$, we define the following length scale
\begin{equation}
\label{eq:W}
\xi \sim \Gamma/W_* \ .
\end{equation}
The length scale $\xi$ represents the size of the region around the crack tip where the
stress/strain concentration is wiped out and as such may be regarded as the crack tip ``load-transfer'' length, i.e.~a length near the crack tip where a characteristic load is transferred to failure processes from the global mechanical fields. It makes no reference to the nature of the reversible elastic deformation that precedes dissipation, e.g.~whether it is predominantly linear or strongly nonlinear, and hence can be rather generally applied to a broad class of materials. It is important to note that in principle there exists a procedure to measure $W_*$; that is, apply uniaxial tension to an as-formed sample in the absence of macroscopic cracks, calculate the area under the resulting force-displacement curve up to catastrophic failure and finally divide the result by the sample's volume. The length scales $\xi$ and $\ell$ are illustrated in {\bf Figure~\ref{fig:2length}a}.

It would be instructive to estimate $\xi$ for a few representative materials. For brittle solids, $W_*$ may be estimated as the linear elastic energy density at the onset of dissipation (after which catastrophic failure typically proceeds), which usually requires a threshold stress. Consider then ideally-brittle materials for which the threshold stress is estimated by Orowan's theoretical strength $\sigma_m\!\sim\!\sqrt{\gamma E/a_0}$ (recall that $a_0$ is an atomistic bond length). Using the latter, we obtain $W_*\!\sim\!\sigma_m^2/E=\gamma/a_0$, which together with $\Gamma\!\sim\!\gamma$ leads to $\xi\!\sim\! a_0$. This shows that $\xi$ is indeed atomistic in the ideally-brittle limit. For glassy polymers like Polymethyl methacrylate (PMMA)~\cite{berry1964fracture} and Polystyrene (PS)~\cite{berry1961fracture}, the threshold stress can be estimated as the crazing stress $\sigma_c$ (or an effective yield stress), which is typically smaller than the ideal strength $\sigma_m$. In addition, for such brittle materials, we typically have $\Gamma\!\gg\!\gamma$, which together with $\sigma_c\!\ll\!\sigma_m$, we obtain $\xi\!\sim\!\Gamma E/\sigma_c^2\!\gg\!a_0$. Using $E\=2.9$ GPa, $\Gamma\=160$ J/m$^2$~\cite{berry1964fracture} and $\sigma_c\!\approx\!70$ MPa for PMMA at room temperature~\cite{Brown1991}, we obtain $\xi\!\sim\!100$ $\mu$m. That is, in this case $\xi$ is indeed significantly larger than atomistic scales.

For ductile (or elasto-plastic) solids $W_*$ is directly measurable and is sometimes known as the ``modulus of toughness''~\cite{DeSilva2013} or, for soft materials, the ``work of extension"~\cite{Li2017a, Takahashi2018}. To estimate it, consider an elastic-perfectly-plastic solid with a yield stress $\sigma_y$ and a failure strain of $\epsilon_f$. The crucial difference compared to the glassy polymers case is that the deformation is no longer predominantly linear elastic prior to the onset of dissipation. That is, the failure strain is much larger than the yield strain $\epsilon_y\!\equiv\!\sigma_y/E$, $\epsilon_f\!\gg\!\epsilon_y$. Consequently, we have $W_*\!\sim\!\sigma_y\epsilon_f$, which implies $\xi\!\sim\!\Gamma/(\sigma_y\epsilon_f)$. For steel, for example, we typically have $\Gamma\!\approx\!10^4$ J/m$^2$, $\sigma_y\!\approx\!0.35$ GPa and $\epsilon_f\!\approx\!10\%$, which leads to $\xi\!\approx\!0.3$ mm. It is important to note that this value of $\xi$ is significantly smaller than the size of the plastic zone, which corresponds to the onset of plastic deformation. A rough estimate for the latter in steel is obtained by noting that $E\!\approx\!200$ GPa and by replacing $\sigma_c$ in the expression for $\xi$ in brittle polymers by $\sigma_y$, i.e.~$\Gamma E/\sigma_y^2\!\approx\!16$ mm, which is indeed much larger than $\xi\!\approx\!0.3$ mm.

The intrinsic material length scale $\xi$ may also be related to size-dependent failure or defect/flaw-sensitivity~\cite{Bazant1997}. In LEFM, where the tensile strength is given by $\sqrt{\Gamma/E c}$, material strength is sensitive to the size of the crack $c$ that is assumed to be much larger than $\xi$, $c\!\gg\!\xi$. However, if $c\!\ll\!\xi$, the stress concentration due to the crack is wiped out by dissipative processes on scale $\xi$ (or on the scale of the plastic zone) and hence the tensile strength may become insensitive to the size of the crack. The sensitivity of failure to defects/flaws size is yet another important distinction between brittle and ductile solids, as $\xi$ (or the scale of the plastic zone) can become quite large for the latter. The concept of flaw-sensitive or flaw-insensitive failure was first proposed for soft materials in~\cite{Chen2017} and then further elaborated on recently in~\cite{Yang2019}, where $\xi$ was termed the ``fracto-cohesive'' length. This concept is further illustrated in {\bf Figure~\ref{fig:2length}b}.

It is important to note that Equations~\ref{eq:ell}-\ref{eq:W} offer scaling estimates for the length scales $\ell$ and $\xi$ respectively, but they do not imply that the not-specified prefactors are necessarily of order unity. Whenever more quantitative estimates are of interest, additional and more accurate considerations of the dimensionless prefactors are required. The two length scales $\ell$ and $\xi$, and the relations between them and the extrinsic/geometric length scales $c$ and $L$, allow one to classify the behavior of a very large class of materials. One can roughly discuss four different classes of materials in this context. First, materials for which both $\xi$ and $\ell$ are typically much smaller than the crack size $c$ (here and in the subsequent discussion $c\!\ll\!L$ is assumed), $\xi\!\sim\!\ell\!\ll\!c$, are stiff brittle materials such as silica glass: $\xi\!\sim\!1$ nm and $\ell\!\sim\!0.1$ nm. Second, materials for which $\xi$ is larger than $\ell$ and comparable to $c$, $\ell\!\ll\!\xi\!\sim\!c$, are stiff ductile materials such as steel: $\xi\!\sim\!0.3$ mm and $\ell\!\sim\!50$ nm. Third, materials for which $\ell$ is significantly larger than $\xi$, but both are significantly smaller than $c$, $\xi\!\ll\!\ell\!\ll\!c$, are soft brittle materials such as brittle hydrogels. Finally, materials for which $\xi$ is smaller than $\ell$ and comparable to $c$, $\ell\!\gg\!\xi\!\sim\!c$, are soft ductile materials such as DN gels~\cite{Gong2010}. The relation $\ell\!\gg\!\xi$ assumed for soft materials, brittle or ductile, is justified by the fact that soft materials can typically sustain a large strain at failure which implies $W_*\!\gg\!E$~\cite{Chen2017}.

The focus of this review paper is on the last two classes of soft materials. For both classes, $\ell$ is a macroscopic length that takes values over a wide range, including $\ell\!\sim\!0.01\!-\!0.1$ mm for Agar gels~\cite{Lefranc2014, Long2016a}, $\ell\!\sim\!1\!-\!10$ mm for Tetra-PEG gels~\cite{Akagi2013}, $\ell\!\sim\!1$ mm for multi-network (MN) elastomers~\cite{Ducrot2014}, $\ell\!\sim\!1\!-\!10$ mm for DN gels~\cite{Nakajima2013} or vulcanized natural rubber~\cite{Rivlin1953} and $\ell\!\sim\!10\!-\!100$ mm for the polyacrylamide/alginate hybrid gels~\cite{Sun2012} or bilayer hybrid gels~\cite{Haque2010,Haque2011}. The macroscopic values attained by $\ell$ imply that nonlinear effects must be accounted for when analyzing crack tip fields in such materials, as elaborated on in Section~\ref{sec:asymptotic}. The length scale $\xi$ spans a huge range, from microscopic to macroscopic values, across different brittle and ductile soft materials. Since $W_*$ is not simply measurable for brittle elastomers and gels due to the sensitivity to pre-existing defects, we estimate in Section~\ref{subsec:tip_dissipation} $\xi$ based on the Lake-Thomas theory~\cite{Lake1967}, which yields $\xi\!\sim\!10$ nm (e.g.~for Agar gels and Tetra-PEG gels). In contrast, soft and very tough materials exhibit much larger $\xi$, e.g., $\xi\!\sim\!0.01\!-\!1$ mm for DN gels~\cite{Gong2010, Nakajima2013, Ahmed2014}, $\xi\!\sim\!0.1-1$ mm for vulcanized natural rubber~\cite{Thomas1958, Greensmith1960} or MN elastomers~\cite{Ducrot2014, Millereau2018}, and $\xi\!\sim\!1\!-\!10$ mm for polyacrylamide/alginate hybrid gels~\cite{Sun2012} or bilayer hybrid gels~\cite{Haque2010,Haque2011}. It is worth noting that in vulcanized natural rubber under cyclic loading, Thomas~\cite{Thomas1958} found that $\xi$ is correlated with the roughness of the newly formed surface due to crack growth, which is consistent with the interpretation of $\xi$ as the size of failure zone. We summarize our rough estimates of $\xi$ and $\ell$ for various materials, with a focus on soft ones, in {\bf Figure~\ref{fig:2length}c}.

\section{Nonlinear elastic crack tip solutions in highly deformable soft materials}
\label{sec:asymptotic}

Our goal here is to discuss the physics on the nonlinear elastic length scale $\ell$, with a focus on key properties of asymptotic nonlinear elastic crack tip solutions, first neglecting the dissipation length scale $\xi$. Due to the nontechnical nature of this review article, we do not provide here full details of the mathematical formulations and solutions, which can be found in a recent review~\cite{Long2015} and the references therein. In order to account for the large elastic deformation on a scale $\ell$, one needs to go significantly beyond linear elasticity (LEFM) in two major respects. First, in the context of LEFM we make no distinction between the reference (undeformed) configuration and the deformed one. While these two configurations are obviously distinct, the geometric differences between them appear only to nonlinear orders and hence are neglected in LEFM. When the deformation is large, these geometric nonlinearities should be taken into account. This feature of highly deformable soft materials introduces significant complications into the problem, because the stress-balance equations (and in general the laws of nature) are formulated in the deformed --- yet a priori unknown --- configuration.

To see this, consider a point whose Cartesian position vector is $\bm{X}\=(X_1,X_2,X_3)$ in the stress-free (undeformed) reference configuration in the presence of a crack. When external driving forces are applied to the cracked body, a point $\bm{X}$ is mapped to a point $\bm{x}=(x_1,x_2,x_3)$ in the deformed configuration, according to $\bm{x}\!=\!\bm{\varphi}(\bm{X})$. The vectorial function $\bm{\varphi}(\bm{X})$, which in general depends on time, is continuous everywhere, except along the crack faces, where it experiences a jump discontinuity. The deformation of material line elements, from $d\bm{X}$ to $d\bm{x}$, is determined by the deformation gradient tensor $\bm{F}(\bm{X})\!=\!\bm{I}+{\B \nabla}{\B u}$ ($\bm{I}$ is the identity tensor), whose Cartesian components are given by $F_{ij}\!=\!\partial{\varphi_i}(\bm{X})/\partial{X_j}$. Since the transformation from $d\bm{X}$ to $d\bm{x}$ includes also rotations which cannot change the physical state of the material element, a proper rotationally invariant measure of deformation is given by the right Cauchy Green tensor $\bm{F}^{T}\bm{F}$. The latter is intrinsically nonlinear in terms of the displacement gradient ${\B \nabla}{\B u}$, which is the geometric nonlinearity that is neglected in LEFM.

In addition to the geometric nonlinearity encapsulated in the deformation measure $\bm{F}^{T}\!\bm{F}$, one should also account for constitutive nonlinearities, i.e.~for a dependence of the strain energy functional on the deformation measure that is stronger than quadratic (LEFM corresponds to linearizing $\bm{F}^{T}\bm{F}$ in terms of ${\B \nabla}{\B u}$ and truncating the strain energy functional to quadratic order in the result). In the following, we consider a rather broad class of highly deformable soft materials, defined by the strain energy functional~\cite{Knowles1977}
\begin{equation}
\label{eq:gnm}
W= \frac{\mu}{2b}\left[\left(1+\frac{b}{n}\left[\tr\!\left(\!\bm{F}^{T}\!\bm{F}\!\right)-3\right]\right)^{\!\displaystyle n}-1\right] \ ,
\end{equation}
where $\mu$ is the linear shear modulus. $W$ in Equation~\ref{eq:gnm}, which is the elastic energy per unit volume in the reference configuration, is known as the generalized incompressible neo-Hookean model (GNH)~\cite{Knowles1977}. Incompressibility, formally expressed as $\det\bm{F}\!=\!1$, determines the other linear elastic constant (say Poisson's ratio, $\nu\=\tfrac{1}{2}$, which implies $\mu\=\tfrac{1}{3}E$). The dimensionless constants $b\!>\!0$ and $n\!>\!\tfrac{1}{2}$ control basic physical properties of the material. In particular, $b$ controls the extent of linear behavior at small deformation and $n$ controls the degrees of strain softening/stiffening, i.e., the change in the tangential modulus at larger deformation. This class of GNH models represents a wide range of nonlinear elastic behaviors in soft polymers, ranging from strain softening ($\tfrac{1}{2}\!<\!n\!<\!1$) to strain stiffening ($n\!>\!1$). The special case of $n\=1$ recovers the celebrated incompressible neo-Hookean model (ideal rubber)~\cite{Holzapfel2000}.

\begin{figure}[ht!]
\begin{centering}
\includegraphics[width=0.9\textwidth]{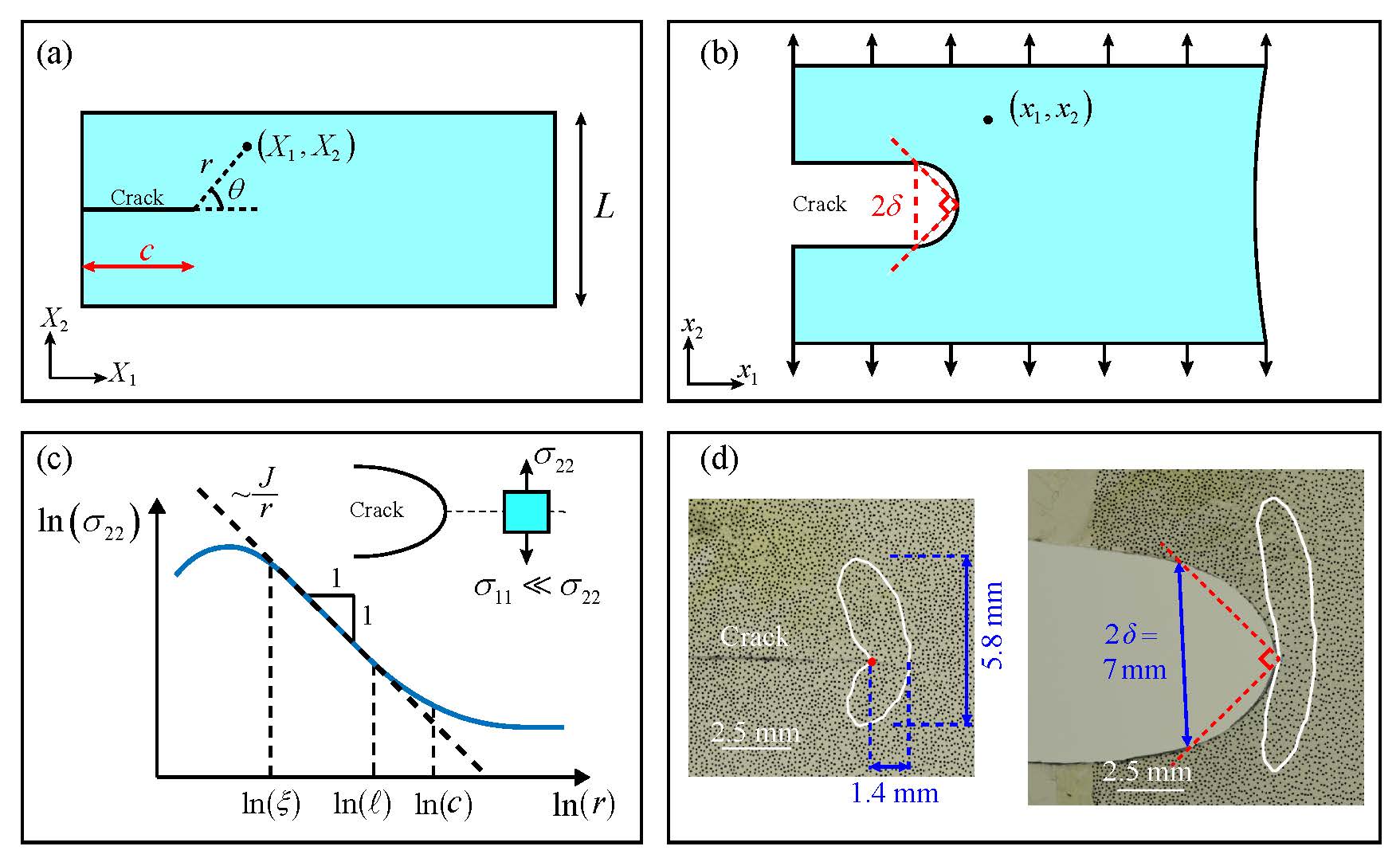}
\par\end{centering}
\caption{Nonlinear elastic crack tip fields. (a) The reference (undeformed) configuration $(X_1,X_2)$ is shown along with a polar coordinate system $(r,\theta)$ centered at the crack tip. (b) The deformed configuration $(x_1,x_2$), corresponding to the reference (undeformed) configuration of panel (a), is shown under tension (denoted by the outgoing arrows at the lower and upper boundaries). The CTOD $2\delta$ is defined as the crack opening spanned by two symmetric rays originating from the crack tip with $90^o$ between them. (c) A schematic representation of the nonlinear elastic tensile (opening) stress $\sigma_{22}$ versus the distance $r$ (in the reference configuration) along the crack line is shown in a log-log scale. For distances larger than the crack length $c$, $r\!\gg\!c$, $\sigma_{22}$ is controlled by the far-field loading. As the tip is approached, $\sigma_{22}$ is amplified due to the presence of the crack. Below the nonlinear elastic length $\ell$, $r\!<\!\ell$, the stress is dominated by the asymptotic singular strongly nonlinear solution of Equation~\ref{eq:s22}, $\sigma_{22}\!\sim\!J/r$ (dashed line). At distances smaller than $\xi$, $r\!<\!\xi$, the singularity is regularized. (inset, top right) The asymptotic singular fields in quasi-static nonlinear elasticity predicts uniaxial stress conditions ahead of the crack tip, $\sigma_{22}\!\gg\!\sigma_{11},\sigma_{12}$. Compare the sketch in panel (c) to that of Figure~\ref{fig:LEFM}c. (d) The nonlinear elastic crack tip zone (white lines, shown in both the reference and deformed configurations) and the CTOD at the initiation of crack growth experimentally measured in a silicone elastomer, see details in~\cite{Qi2019}. \label{fig:asympt_fields}}
\end{figure}

Obtaining analytic crack tip field solutions in the framework of 3D nonlinear elasticity is practically impossible, hence we focus here on solutions that feature 2D symmetry. In particular, if a tensile (opening) stress is applied along the $X_2$ axis and the crack is located along the $X_1$ axis, cf.~{\bf Figure~\ref{fig:asympt_fields}a-b}, we consider solutions that are independent of the out-of-plane coordinate $X_3$ (more formally, we consider a mode-I plane-stress problem, which corresponds to thin samples such that all of the stress components perpendicular to the $X_1\!-\!X_2$ plane vanishes, $\sigma_{3i}\=0$ for $i\=1,2,3$). Assigning a polar coordinate system $(r,\theta)$ to the crack tip in the reference configuration, cf.~{\bf Figure~\ref{fig:asympt_fields}a}, and transforming the quasi-static mechanical balance equations and boundary conditions for the in-plane true (Cauchy) tensor $\B \sigma(\B x)$ (the force per unit area in the deformed configuration $\B x$) into the reference (undeformed) configuration, the asymptotic solution in the $r\!\to\!0$ limit (i.e.~neglecting the dissipation length $\xi$) can be obtained~\cite{Geubelle1994, Long2011}. A crack is physically defined as composed of surfaces that cannot sustain stress, and obtaining the asymptotic solution crucially depends on the fact that the stress (traction) free boundary conditions in the deformed configuration $\B x$ transform into stress (traction) free boundary conditions along the crack surfaces ($\theta\=\pm\pi$) in the reference configuration.

In the strongly nonlinear limit as the crack tip is approached, $r\!\to\!0$, the leading order solution for the in-plane true (Cauchy) tensor $\B \sigma$, in terms of the reference coordinate system $(r,\theta)$, takes the form~\cite{Long2015}
\begin{equation}
\label{eq:s22}
\sigma_{22}(r\!\to\!0,\theta) = \frac{J}{r}h(\theta; n), \qquad\qquad \frac{\sigma_{12}(r\!\to\!0,\theta)}{\sigma_{22}(r\!\to\!0,\theta)}\to0, \qquad\qquad
\frac{\sigma_{11}(r\!\to\!0,\theta)}{\sigma_{22}(r\!\to\!0,\theta)}\to 0 \ ,
\end{equation}
where $h(\theta; n)\=\pi^{-1}\Big[\sqrt{1-\left(\tfrac{n-1}{n}\right)^{2}\sin^{2}\!\theta}-\left(\tfrac{n-1}{n}\right)\cos\theta\Big]$~\cite{Long2015}. Here $J$ denotes the value of the so-called path-independent J-integral~\cite{Knowles1973, Geubelle1994}, which depends on the loading configuration of the global (not asymptotic) crack problem and plays the role of the stress intensity factor $K$ in the LEFM asymptotic solution. In fact, $J$ equals the energy release rate $G$, discussed above in the context of LEFM (though it is a more general concept applicable to any elastic strain energy functional). The strongly nonlinear elastic solution in Equation~\ref{eq:s22}, which is illustrated in {\bf Figure~\ref{fig:asympt_fields}c}, is markedly and qualitatively different from the corresponding LEFM solution (cf.~{\bf Figure~\ref{fig:LEFM}}). In the latter, all of the components of $\bm{\sigma}$ feature the same singularity $\sim\!1/\sqrt{r}$, and in fact $\sigma_{22}(r\!\to\!0,\theta\=0)\=\sigma_{11}(r\!\to\!0,\theta\=0)$ in the quasi-static LEFM solution~\cite{Zehnder2012}, cf.~{\bf Figure~\ref{fig:LEFM}c}. In the strongly nonlinear elastic solution of Equation~\ref{eq:s22}, the tensile component $\sigma_{22}$ completely dominates and features a stronger singularity $\sim\!1/r$. Consequently, the state of elastic stress near crack tips in highly deformable soft materials is predominantly of uniaxial tension nature, while in linear elastic materials is predominantly biaxial in nature under plane-stress conditions. We note in passing that the true stress field in Equation~\ref{eq:s22} is expressed using the reference polar coordinates, where it attains a separable form in terms of $r$ and $\theta$. In the deformed configuration, however, the stress fields are no longer separable in terms of the deformed polar coordinates~\cite{Long2015}.

The strain energy per unit volume in the undeformed configuration, $W$ of Equation~\ref{eq:gnm}, that corresponds to the strongly nonlinear solution of Equation~\ref{eq:s22} takes the asymptotic form
\begin{equation}
\label{eq:wtip}
W(r\!\to\!0,\theta) =\frac{J}{2\,n\,r}h(\theta; n) \ .
\end{equation}
Note that the scaling $W\!\sim\!J/r$ is also found in LEFM, where $W\!\sim\!\sigma^2/E\!\sim\!(K/\sqrt{r})^2/E\!\sim\!G/r\!\sim\!J/r$ ($J\=G$ was used). Yet, as explained above, the scaling of the tensile stress component $\sigma_{22}$ is different, which can be employed to define a nonlinear elastic length scale. Since the strongly nonlinear elastic prediction of $\sigma_{22}\!\sim\!J/r$ is scaling-wise identical to the prediction of the weakly nonlinear theory~\cite{Bouchbinder2014,Livne2010,Bouchbinder2009, Livne.08,Bouchbinder.08} discussed in Section~\ref{sec:2lengths}, comparing it to the LEFM solution yields the same crossover length $\ell\!\sim\!J/E\=G/E\=\Gamma/E$, as in Equation~\ref{eq:ell}. Obviously, the prefactors in the relation $\ell\!\sim\!\Gamma/E$ are different, where the prefactor in the strongly nonlinear elastic case is smaller (simply because the strongly nonlinear elastic zone resides inside the weakly nonlinear one). Finally, we would like to stress again, as done in Section~\ref{sec:2lengths}, that the scaling relation $\ell\!\sim\!\Gamma/E$ has been obtained, either using the weakly or strongly nonlinear solutions, by comparing these to the LEFM solution. Yet, when considering highly deformable soft materials, there are situations in which the LEFM asymptotic solution has no range of validity at all. Even in such situations, the scaling relation $\ell\!\sim\!\Gamma/E$ might be useful, as will be discussed next (and later on in the article).

Are there other relevant and measurable physical quantities that are related to the length scale $\ell\!\sim\!\Gamma/E$? To address this question, let us consider the crack tip opening displacement (CTOD), which was already invoked above in relation to LEFM and is often also employed in elastic-plastic fracture mechanics~\cite{Rice1968, Shih1981}. As shown in {\bf Figure~\ref{fig:asympt_fields}b}, the CTOD --- denoted as $2\delta$ --- is the opening displacement spanned by two symmetric rays originating from the crack tip with a $90^o$ angle between them.  In the framework of LEFM, where the local crack opening profile is parabolic, it can be shown that
\begin{equation}
\label{eq:lefmctod}
 \delta \sim \Gamma/E \sim \ell \ ,
\end{equation}
which shows that the CTOD scales with the nonlinear elastic length $\ell$. Does this scaling persist for strongly nonlinear elastic crack tip solutions? For the GNH models discussed above, the asymptotic solutions the CTOD depend on the exponent $n$~\cite{Geubelle1994, Long2011} and take the form~\cite{Long2015}
\begin{equation}
\label{eq:nlctod1}
\delta \sim \left(J/E\right)^{\alpha(n)} = \left(\Gamma/E\right)^{\alpha(n)} \ ,
\end{equation}
where the function $\alpha(n)$ varies between $\sim\!0.8$ and $\sim\!1.1$. We thus conclude that the relation $\delta\!\sim\!\ell$ {\em approximately} holds also for a wide class of highly deformable soft materials. The length $\delta$, or equivalently $\ell$, is an important physical quantity that characterizes the geometry of crack tips. In particular, it can quantify the extent of crack blunting that can be sustained before crack initiation~\cite{Hui2003, MacDonald2020}.

To demonstrate the practical relevance of the nonlinear elastic length $\ell$, we refer to the recent experimental work of~\cite{Qi2019}, where the deformation fields around a tensile crack in a soft silicone elastomer were measured by a particle tracking method, cf.~{\bf Figure~\ref{fig:asympt_fields}d}. The region of dominance for the nonlinear crack tip fields upon crack initiation was found to exhibit a butterfly shape, with a length of $\sim\!1.4$ mm directly ahead of the crack tip and a width of $\sim\!6$ mm perpendicularly to the crack line. The half CTOD $\delta$ at crack initiation was estimated to be $\sim\!3.5$ mm. On the other hand, using the experimentally determined values of $\Gamma\=120$ J/m$^2$ at crack initiation and $E\=3\mu\=60$ kPa, we obtain that $\ell\!\sim\!2$ mm, in quantitative agreement with both the nonlinear zone size and the half CTOD $\delta$.

Finally, we note that Equation~\ref{eq:wtip} predicts a singular strain energy density $W$ as the crack tip is approached, $r\!\to\!0$. This singular behavior cannot persist to indefinitely small scales, but is rather cutoff at the dissipation scale $\xi$. As discussed in Section~\ref{sec:2lengths}, $W$ cannot exceed the critical energy density $W_*$ for failure, and therefore Equation~\ref{eq:wtip} breaks down at $r\=\xi$ according to $J/\xi\!\sim\!W_*$. The latter, together with $J\=\Gamma$, recovers Equation~\ref{eq:W}, which shows that $\xi$ indeed corresponds to the scale of the failure zone around the crack tip, as will be further discussed in Section~\ref{subsec:tip_dissipation}.

\subsection{Inertial effects during dynamic crack propagation}
\label{subsec:dynamic}

The discussion up to now focused on cracks at the onset of propagation, i.e.~on fracture initiation, under quasi-static loading conditions. Once a crack is set into motion, many new physical effects associated with its propagation velocity $v$ may emerge. When the material of interest features bulk rate-dependence (i.e.~strain-rate sensitivity), interesting physical effects emerge, as will be discussed in Section~\ref{subsec:visco}. Here we would like to briefly discuss physical situations in which material inertia plays important roles in crack propagation, i.e.~when $v$ is comparable to an elastic wave-speed, say the shear wave-speed $c_s$. We exclude from the discussion material rate-dependence in the bulk, but allow the fracture energy $\Gamma$ to depend on the crack propagation velocity $v$. That is, in general we have~$\Gamma(v)$, which is rather generically a mildly increasing function of $v$, and in particular, $\Gamma(v)\!>\!\Gamma(v\!\to\!0)$~\cite{Livne2010}.

Under these conditions, the main concepts and physical quantities discussed in previous sections, in particular the intrinsic length scales $\xi$ and $\ell$, remain valid. The strongly dynamic conditions we consider here, i.e.~$v\!\sim\!c_s$, have important quantitative and quantitative implications. The nonlinear length scale $\ell(v)\!\sim\!\Gamma(v)/E$ maintains the scaling structure of Equation~\ref{eq:ell}, but the prefactor in this scaling relation is significantly larger in the strongly dynamic regime than in the initiation/quasi-static regime~\cite{Bouchbinder2014,Livne2010,Bouchbinder.09b}. That is, $\ell(v/c_s\!\sim\!1)\!\gg\!\ell(v/c_s\!\ll\!1)$ mainly due to the $v$-dependence of the prefactor (and not due to the mild variation of $\Gamma(v)$ with $v$). This implies that even for brittle soft materials, such as polyacrylamide hydrogels~\cite{Bouchbinder2014,Livne2010,Livne.08, Livne2007}, which break under small far-field strains for which the assumptions of LEFM are valid, a non-negligible nonlinear zone of size $\ell(v)$ might develop around strongly dynamic crack tips.

\begin{figure}[ht!]
\begin{centering}
\includegraphics[width=\textwidth]{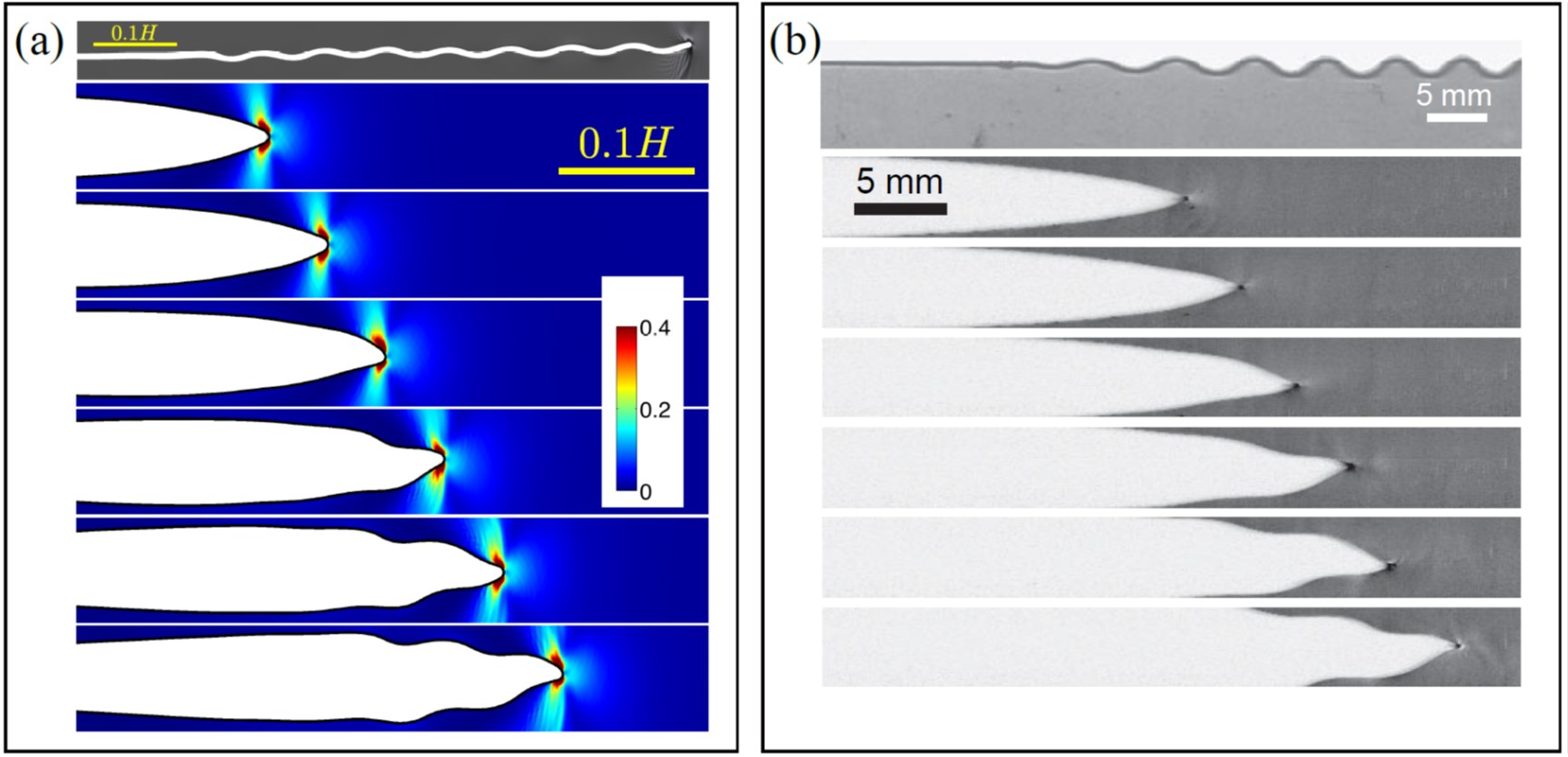}
\par\end{centering}
\caption{(a) The theoretical prediction of a high-speed 2D oscillatory instability, using large-scale computer simulations~\cite{Chen2017a}, based on near crack tip nonlinearity in brittle materials. The top part shows the crack trajectory and the lower part exhibits a series of snapshot revealing the onset of instability in the deformed configuration. The color code corresponds to the normalized strain energy density $W/\mu$, where $W$ corresponds to Equation~\ref{eq:gnm} with $n\!=\!1$, i.e.~to incompressible neo-Hookean materials. The strong amplification of deformation near the propagating tip is evident. The oscillation wavelength has been shown to scale linearly with the nonlinear elastic length $\ell(v)$. The scale bar corresponds to $0.1H$, where $H$ is the height of the system, highlighting the fact that the instability is controlled by the intrinsic length scale $\ell(v)$ and not by extrinsic/geometric length scales (such as $H$). (b) The corresponding experimental observation in a polyacrylamide hydrogel~\cite{Livne2007}. Note that the wavelength is in the mm range, while the quasi-static $\ell(v\!\to\!0)\!\approx\!100\,\mu$m. This difference is accounted for by the increase of $\ell(v)$ under strongly dynamic conditions, see text for additional discussion. For a quantitative comparison between the theory and the experiments, see~\cite{Chen2017a}. \label{fig:oscillations}}
\end{figure}

The most striking effect of the nonlinear zone of size $\ell(v)$ in strongly dynamic situations is in fact not just quantitative, but rather qualitative. Most notably, it has been recently shown experimentally~\cite{Bouchbinder2014,Livne2007,Goldman2012} and theoretically~\cite{Bouchbinder.09b,Chen2017a, Lubomirsky2018} that $\ell(v)$ controls a spontaneously symmetry-breaking instability in dynamic fracture of brittle materials. In particular, it has been shown that straight cracks propagating in 2D brittle materials that feature a nonlinear zone near their tip lose their stability upon surpassing a high-speed threshold (close to $c_s$) and start to oscillate/wiggle, see {\bf Figure~\ref{fig:oscillations}}. The wavelength of oscillations has been shown to scale linearly with $\ell(v)$~\cite{Goldman2012, Lubomirsky2018}, thus demonstrating that the nonlinear length $\ell(v)$ --- absent in LEFM --- plays a decisive role in dynamic fracture instabilities in brittle materials.

A length scale related to $\ell(v)$ has also been demonstrated recently in soft ductile/tough DN gels~\cite{Kolvin2018}, which can undergo dynamic fracture and hence feature strong inertial effects. By analyzing the experimentally observed CTOD in dynamic cracks in DN gels, which significantly deviates from the parabolic CTOD of LEFM, a $v$-dependent length scale has been extracted~\cite{Kolvin2018}. Inspired by the relation $\ell(v)\!\sim\!\Gamma(v)/E$, it was further shown that the CTOD-extracted length is proportional to the stored elastic energy (which is balanced by $\Gamma(v)$) and significantly increases as $v\!\to\!c_s$, all consistent with the properties of $\ell(v)$. Finally, we note that under strongly dynamic conditions, inertial effects also lead to an increase in the dissipation length $\xi$~\cite{Kolvin2018}.

\section{Dissipative processes in highly deformable soft materials}
\label{sec:dissipation}

The discussion in the previous section focused on near crack tip nonlinear elastic deformation in soft materials and on the associated nonlinear elastic length scale $\ell$. Here we shift our focus to dissipative processes in highly deformable soft materials, especially to the dissipation length scale $\xi$ and to the fracture energy $\Gamma$, which quantifies the dissipation associated with cracks. It would be instructive to distinguish between two distinct contributions to $\Gamma$. The first contribution is the intrinsic fracture energy $\Gamma_0$, which represents crack tip associated dissipation directly related to the material separation process inherent in fracture. The second contribution to the fracture energy, $\Gamma_{\rm D}$, represents bulk dissipation that takes place further away from the crack tip in regions where material load-bearing has not been lost. Consequently, we introduce the decomposition
\begin{equation}
\Gamma=\Gamma_0 + \Gamma_{\rm D} \ ,
\label{eq:decomposition}
\end{equation}
which will be useful for the discussion below.

It is important to note that the length scales associated with $\Gamma_0$ and $\Gamma_{\rm D}$ do not in general coincide with the length scale $\xi$. Indeed, the definition of $\xi$ involves in addition to $\Gamma$ also the work of extension $W_*$, and hence there is no general one-to-one mapping between $\xi$ and the length scale that is associated with $\Gamma$. The length scale $\xi$ characterizes the size of the region around the crack tip where the stress/strain concentration is wiped out. That is, $\xi$ may be regarded as the crack tip ``load-transfer'' length, i.e.~a length near the crack tip where a characteristic load is transferred to failure processes from the global mechanical fields. These issues will be further discussed below through more explicit examples.

The bulk dissipation contribution to the fracture energy, $\Gamma_{\rm D}$, may involve both rate-independent and rate-dependent dissipative processes, and can also be strongly coupled to processes contributing to $\Gamma_0$. The interaction of these processes with the spatial structure of the near crack tip fields give rise to non-trivial effects. Rate-independent dissipative processes typically depend on the magnitude of the stress/strain, which in turn depend on the distance from the crack tip, where the mechanical fields are strongly amplified. Consequently, different physical processes may be activated at different spatial positions, depending on their distance from the tip. Rate-dependent dissipative processes, which depend on the magnitude of the local strain-rate, give rise to even more intricate physical effects. To see this, consider a crack steadily propagating at a constant speed $v$. That means that the near tip inhomogeneous stress/strain fields are dragged with the crack tip as a constant speed, implying that different regions in the material experience different strain-rate. In particular, regions close to the crack tip experience larger strain-rates than regions further away, and hence their physical response is different in rate-dependent materials.

In Subsection~\ref{subsec:tip_dissipation}, we discuss the intrinsic fracture energy $\Gamma_0$, and the relation between $\xi$ and $\Gamma$ in situations in which $\Gamma_{\rm D}$ is negligibly small. In Subsection~\ref{subsec:Mullins}, we discuss rate-independent bulk dissipation and its contribution to $\Gamma_{\rm D}$. Finally, in Subsection~\ref{subsec:visco}, we discuss rate-dependent bulk dissipation --- with a focus on viscoelasticity ---, mainly highlighting the effect of the crack propagation speed $v$ on the structure of the crack tip fields and on $\Gamma_{\rm D}$.

\subsection{Crack tip associated dissipation}
\label{subsec:tip_dissipation}

\begin{figure}[ht!]
\begin{centering}
\includegraphics[width=0.8\textwidth]{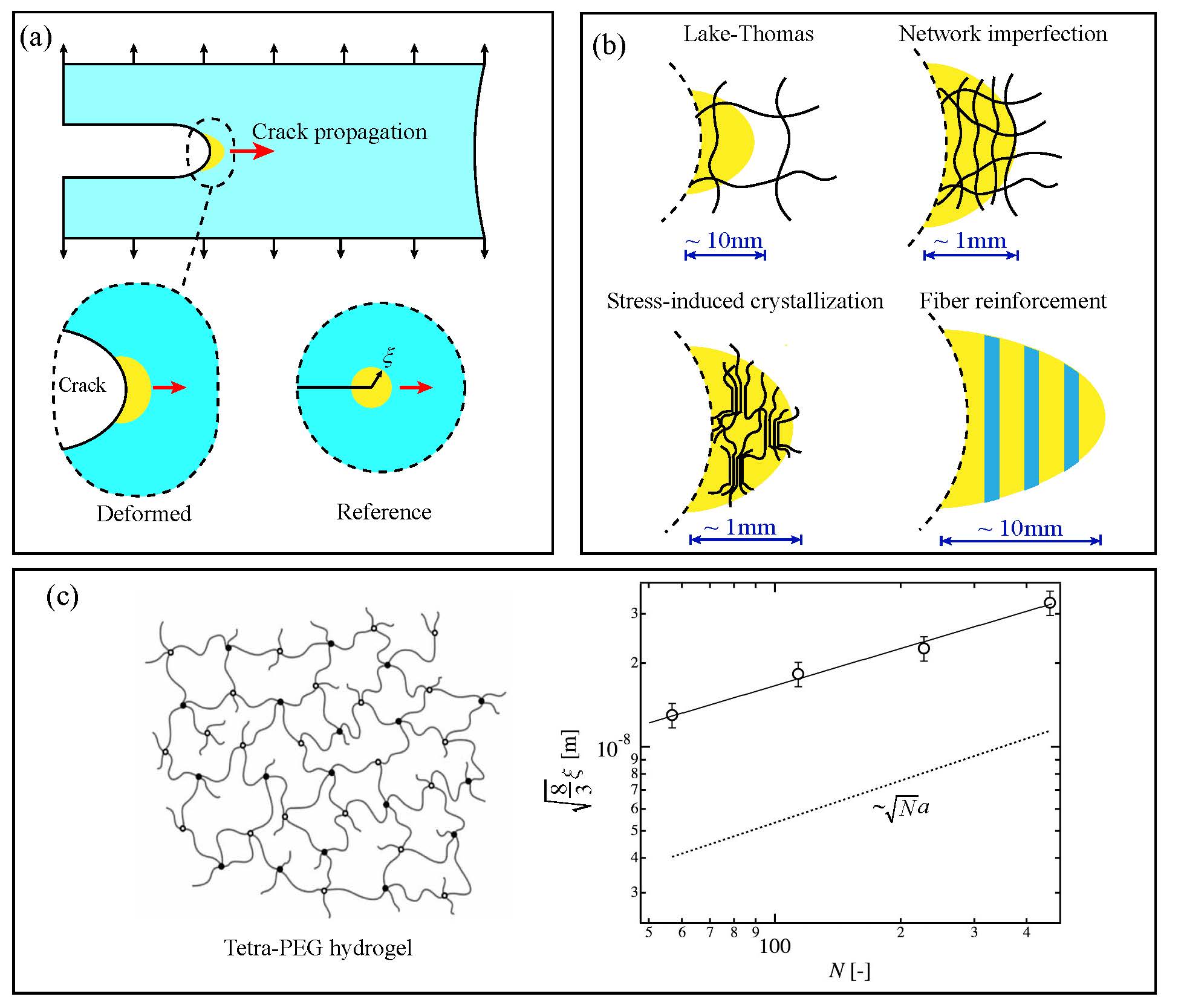}
\par\end{centering}
\caption{(a) A crack steadily propagating in a soft material under far-field tensile loading conditions. Dissipation is assumed to occur only near the crack tip, in the yellow region, such that $\Gamma\!\approx\!\Gamma_0$ (see text for details). Under these conditions, the typical size of the near tip dissipation zone is identified with $\xi$ of Equation~\ref{eq:W}, as shown at the bottom raw (zoom-in on the crack tip region, in both the deformed and reference configurations). (b) Various mesoscopic structures can give rise to widely varying crack tip dissipation lengths, see text for additional details. (c) A schematic sketch of the network structure of Tetra-PEG gels~\cite{Sakai2010} is shown on the left. The experimentally extracted length $\xi$ (multiplied by $\sqrt{8/3}$~\cite{Akagi2013}) in the Lake-Thomas expression of Equation~\ref{eq:LT} vs.~the number of monomers in a chain $N$ is shown on the right (Adapted from Figure 1 of~\cite{Sakai2010}). The solid line corresponds to $\xi\!\sim\!N^{0.45}a$ ($a$ in the monomer length), which is in reasonable agreement with the Lake-Thomas prediction $\xi\!\simeq\!N^{0.5}a$ based on Gaussian chain statistics (dashed line), albeit with an enhanced prefactor~\cite{Akagi2013}. See text for additional discussion. \label{fig:tip_dissipation}}
\end{figure}

We focus in this subsection on the intrinsic fracture energy $\Gamma_0$ and on its relation to the dissipation length $\xi$ when $\Gamma_{\rm D}$ is negligible, i.e.~when $\Gamma\!\approx\!\Gamma_0$. We consider a crack propagating slowly and steadily under far-field tensile loading, see {\bf Figure~\ref{fig:tip_dissipation}a}. Crack tip dissipation occurs in the region schematically marked in yellow in {\bf Figure~\ref{fig:tip_dissipation}a}, and zoomed in on in the lower part (both in the deformed and reference configurations). In the ideal brittle limit, the typical size of the dissipation zone (yellow region) is atomistic, $a_0$, and the intrinsic fracture energy $\Gamma_0$ equals the bare surface energy $2\gamma$, as discussed in Section~\ref{sec:intro}. The bare surface energy is typically $\sim\!1$ J/m$^2$, which is significantly smaller than the experimentally measured values for soft brittle polymers, which are typically in the range of $\sim\!10\!-\!100$ J/m$^2$.

This discrepancy can be associated with the existence of polymeric degrees of freedom that are not taken into account in the ideal brittle picture. In quantitative terms, the discrepancy was addressed by the Lake-Thomas molecular theory~\cite{Lake1967}, which considered a regular (lattice-like) polymeric network. The theory suggests that when a stretched polymer chain ahead of the crack tip breaks, every pair of monomers in the chain loses energy that is comparable to the bond interaction energy $U_b$. Therefore, in this picture $\Gamma\=\Gamma_0$ is not only associated with the energy needed to break a single bond, but is rather multiplied by the number of bonds/monomers $N$ in the polymer chain. Consequently, $\Gamma$ can be expressed as~\cite{Creton2016,Tang2017}
\begin{equation}
\label{eq:LT}
\Gamma \approx \nu_x N U_b\,\xi \ ,
\end{equation}
where $\nu_x$ is the number of chains per unit reference volume and $\xi$ is the length ahead of the tip where chain scission takes place. Using the notation $\xi$ is justified only if Equation~\ref{eq:LT} identifies with Equation~\ref{eq:W}; this is indeed the case because the energy per unit reference volume of the stretched polymeric network at failure is $\nu_x N U_b$, which is --- by definition --- the work of extension $W_*$. Hence, Equation~\ref{eq:LT} can be in fact expressed as $\Gamma\!\approx\!W_* \xi$, which is identical to Equation~\ref{eq:W}. Finally, the Lake-Thomas theory also identifies $\xi$ (in our language) with the network's mesh size, cf.~{\bf Figure~\ref{fig:tip_dissipation}b} (top-left). The latter can be estimated using random walk statistics, which leads to $\xi\!\sim\!\sqrt{N}a$, where $a$ in the monomer length.

The predictions of the Lake-Thomas theory have been recently tested for Tetra-PEG gels, which feature a rather regular network structure (cf.~{\bf Figure~\ref{fig:tip_dissipation}c}), that was carefully and systematically controlled experimentally~\cite{Sakai2010}. By controlling $\nu_x$ and $N$, and by independently measuring $\Gamma$, $\xi$ can be extracted from Equation~\ref{eq:LT} once $U_b$ is estimated~\cite{Akagi2013}. The measured values of $\Gamma$ were in the $10\!-\!50$ J/m$^2$ range and $\xi$ was found to be a multiple of $10$ nm. The $N$-dependence of $\xi$ is presented in {\bf Figure~\ref{fig:tip_dissipation}c}, demonstrating a $\xi\!\sim\!N^{0.45}$ scaling that is reasonably consistent with the prediction $\xi\!\sim\!N^{0.5}a$, which is based on Gaussian (freely jointed) chains. These experimental results quantitatively support the Lake-Thomas theory for gels featuring rather regular polymeric networks, and clearly demonstrate that polymeric degrees of freedom can lead to $\xi\!\gg\!a_0$ and $\Gamma\!\gg\!2\gamma$.

The Lake-Thomas predictions discussed above, obtained for dry polymeric networks, can be extended for swollen polymeric networks in which the network only occupies a fraction $\phi_p$ of the total volume ($0\!<\!\phi_p\!\le\!1$). Swelling decreases the chain density according to $\nu_x\!\to\!\phi_p \nu_x$ and increases the dissipation length (network mesh size in this case) according to $\xi\!\to\!\phi_p^{-1/3}\xi$, assuming affine deformation~\cite{Tang2017}. Consequently, the fracture energy of Equation~\ref{eq:LT} decreases according to $\Gamma\!\to\!\phi_p^{2/3}\Gamma$. In addition, the work of extension transforms according to $W_*\!\to\!\phi_p W_*$, which implies that the dissipation length follows $\xi\!\sim\!\Gamma/W_*\!\sim\!\phi_p^{-1/3}N^{1/2}a$. The scaling of Young's modulus with $\phi_p$ is more complicated and may vary depending on physical conditions of the gel network~\cite{Hoshino2018, sakai2012effect}. Assuming an ideal network consisting of Gaussian chains that are relaxed in the dry state and then swollen, it was shown that $E\!\sim\!\phi_p^{1/3}\nu_x k_B T$~\cite{Yang2019, cai2012equations}, where $k_B$ is Boltzmann's constant and $T$ is the absolute temperature. Adopting this scaling for Young's modulus, the nonlinear elastic length follows $\ell\!\sim\!\Gamma/E\!\sim\!\phi_p^{1/3}N^{3/2}aU_b/k_BT$. Using representative values for dry elastomers: $U_b\=5\!\times\!10^{-19}$ J, $a\=0.5$ nm, $N\=1000$ and $\phi_p\=1$, we find at room temperature $\ell\!\sim\!2$ mm and $\xi\!\sim\!15$ nm. If the same network is swollen according to $\phi_p\!=\!0.1$, then we find $\ell\!\sim\!1$ mm and $\xi\!\sim\!30$ nm.

The Lake-Thomas theory assumes that crack tip dissipation occurs only in a single layer of chains in an idealized network with a uniform chain length. However, since chains are already severely stretched near the crack tip, inhomogeneity in the network may play a critical role in determining which chain would fail first~\cite{Yang2019}. In particular, shorter chains can experience higher forces even if they are further away from the crack tip; consequently, network imperfection can offset the strong stress concentration at the crack tip, significantly enlarging the crack tip dissipation zone and increasing $\Gamma_0$. This effect of network imperfection is illustrated in {\bf Figure~\ref{fig:tip_dissipation}b} (top-right), where the crack tip dissipation zone can be extended to the mm range in polyacrylamide gels~\cite{Yang2019}. It should be emphasized that polyacrylamide gels exhibiting $\xi\!\sim\!1$ mm have a much lower cross-link density than typical brittle polyacrylamide gels, e.g., those referred to in Section~\ref{subsec:dynamic}, where $\xi\!\sim\!20\,\mu$m~\cite{Livne2010}. In addition to network imperfection, mesoscale structures near the crack tip can also significantly increase the crack tip dissipation zone and hence $\Gamma_0$. Examples, as illustrated in {\bf Figure~\ref{fig:tip_dissipation}b} (bottom row), include stress-induced crystallization in natural rubber~\cite{Trabelsi2002, Persson2005}, and fiber reinforcement in polydimethylsiloxane (PDMS) composites~\cite{Wang2019, Xiang2019} or fabric reinforced polyampholyte hydrogel composites~\cite{king2015extremely,huang2019superior,hui2019size}.

\subsection{Rate-independent bulk dissipation: Damage-induced softening and the Mullins effect}
\label{subsec:Mullins}

\begin{figure}[ht!]
\begin{centering}
\includegraphics[width=0.75\textwidth]{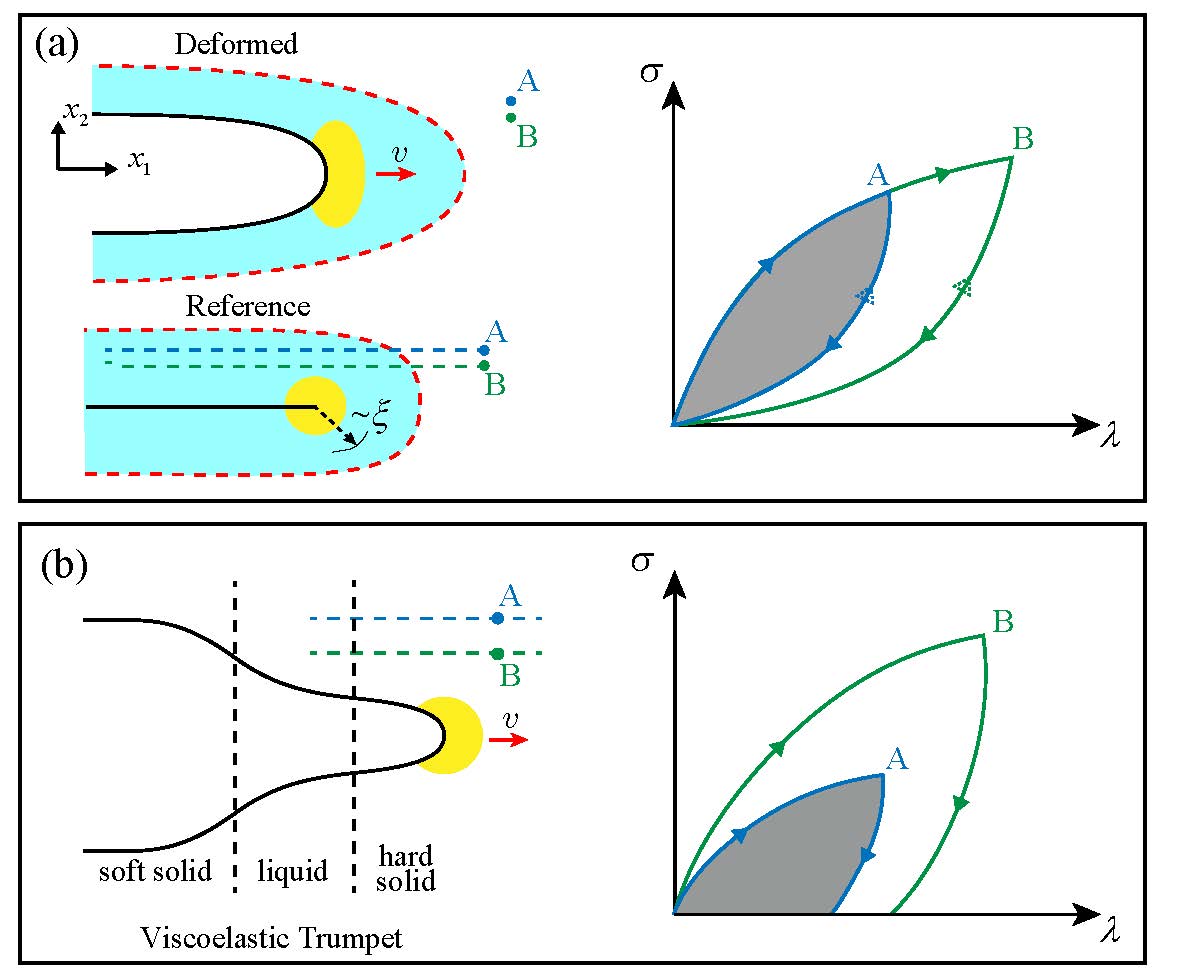}
\par\end{centering}
\caption{(a) A crack steadily propagating at a speed $v$ as in {\bf Figure~\ref{fig:tip_dissipation}a} is shown on the left, but this time for soft materials in which bulk dissipation exists in addition to crack tip dissipation (yellow region as in {\bf Figure~\ref{fig:tip_dissipation}}). The bulk dissipation zone and its wake are enclosed within the red dashed line. Under such conditions, the dissipation length of Equation~\ref{eq:W} (shown schematically at the bottom part) cannot be identified with the size of the crack tip dissipation zone. Its relation to the bulk dissipation zone is discussed in the text. Two loading-unloading hysteresis loops (stress $\sigma$ vs.~stretch $\lambda$) representatives of rate-independent soft materials are shown on the right. They reveal amplitude dependence characteristic of the Mullins effect, see text for additional details and note that the empty arrow heads represent the response upon subsequent reloading. The two hysteresis loops also correspond to the different material response at different locations away from the crack line, marked by A and B on the left (see text for additional discussion). (b) The viscoelastic trumpet structure of cracks in rate-dependent soft materials is shown on the left, see text for details and discussion. On the left two loading-unloading hysteresis loops are shown as in panel (a), but this time for rate-dependent soft materials (see text for discussion). \label{fig:bulk_dissipation}}
\end{figure}

In the previous subsection we discussed several classes of soft brittle materials whose fracture energy $\Gamma$ is dominated by the intrinsic fracture energy $\Gamma_0$. This is not the case for tougher, more ductile, soft materials. Consider, for example, DN gels~\cite{Gong2003, Gong2010}. These gels consist of two interpenetrating networks: a highly cross-linked, swollen and thus stiff network, and a loosely cross-linked and thus extensible network. A similar molecular architecture has been achieved by mixing ionically cross-linked alginate and covalently cross-linked polyacrylamide~\cite{Sun2012}, resulting in an extremely stretchable and tough material, featuring a fracture energy of $\Gamma\!\approx\!9000$ J/m$^2$. Energy dissipation in such materials, and in DN gels in particular, is related to bond breaking in the stiff (first) network~\cite{Gong2010, Ducrot2014, Millereau2018}, which is consequently termed the sacrificial network, while the extensible (second) network maintains the load-bearing capacity of the material. The intrinsic fracture energy $\Gamma_0$ of DN gels is estimated as $\Gamma_0\=1\!-\!10$ J/m$^2$~\cite{Gong2010}, corresponding to dissipation taking place in the near tip region marked in yellow on left part of {\bf Figure~\ref{fig:bulk_dissipation}a}. The typical size of this region can be roughly estimated as the length of fragmented blocks in the stiff network, being in the $0.1\!-\!1\,\mu$m range~\cite{Gong2010}.

The overall fracture energy of DN gels, however, is measured to be $\Gamma\=100\!-\!3000$ J/m$^2$~\cite{Gong2010, Ahmed2014}, which is about two orders of magnitude larger than $\Gamma_0$. This implies that the bulk dissipation contribution $\Gamma_{\rm D}$ in Equation~\ref{eq:decomposition} dominates the fracture energy. The vast difference between $\Gamma_0$ and $\Gamma_{\rm D}$ is also reflected in the associated length scales. The size of the bulk dissipation zone, schematically marked by the red dashed line in {\bf Figure~\ref{fig:bulk_dissipation}a}, is measured in DN gels to be several hundreds $\mu$m~\cite{Gong2010}, which is indeed much larger than the length scale associated with $\Gamma_0$. Plugging $\Gamma\!\approx\!1000$ J/m$^2$ and $W_*\!\approx\!10$ MJ/m$^3$~\cite{Gong2010, Nakajima2013} into Equation~\ref{eq:W}, we obtain $\xi\!\approx\!100\,\mu$m for DN gels. This length is comparable to, yet somewhat smaller than, the size of the bulk dissipation zone in this tough soft material~\cite{Gong2010}. In general, the dissipation length $\xi$ is expected to be smaller than the size of the bulk dissipation zone in soft materials for which $\Gamma_{\rm D}\!\gg\!\Gamma_0$, but to scale with it, as demonstrated in a recent study~\cite{Zhang2015}.

The energy dissipation quantified by $\Gamma_{\rm D}$ is manifested in the hysteresis loop of the stress $\sigma$ versus stretch $\lambda$ curve (or alternatively the stress-strain curve) over a loading-unloading cycle in a uniaxial tensile test, see the blue curve on the right part of {\bf Figure~\ref{fig:bulk_dissipation}a}. Unlike metals, the permanent deformation after a loading-unloading cycle in DN gels is relatively small (it is set to zero in the schematic sketch in {\bf Figure~\ref{fig:bulk_dissipation}a}). Upon reloading, the new loading curve lies almost on top of the unloading curve from the previous loading cycle, as long as the maximum stress in the previous loading is not exceeded. When the maximal stress is exceeded, the unloading curve features additional softening (i.e.~it approaches the unloaded state with a smaller slope), see the green curve on the right part of {\bf Figure~\ref{fig:bulk_dissipation}a}. This phenomenon --- analogous to the Mullins effect in filled rubbers~\cite{Mullins1969, Diani2009} --- indicates that softening and dissipation in the DN gels is due to partial damage of the sacrificial network, which occurs only during loading. Note that the Mullins-like effect depends on the magnitude of the stress/strain, but is mostly independent of the loading/unloading rate.

This amplitude dependence of the hysteresis curve in rate-independent soft materials such as DN gels, which implies also different levels of dissipation (quantified by the area under the hysteresis loop), has consequences for propagating cracks. Most notably, material points located at different heights above the crack line --- for example the points A and B on the left part of {\bf Figure~\ref{fig:bulk_dissipation}a} --- experience different stress-strain curves (cf.~the right part of {\bf Figure~\ref{fig:bulk_dissipation}a}, where A and B are marked). Consequently, analytical treatments of the near tip mechanical fields in these materials is highly involved and remain an open problem~\cite{qi2018fracture}.

\subsection{Rate-dependent bulk dissipation: Viscoelasticity}
\label{subsec:visco}

In the previous subsection we discussed the bulk dissipation contribution to the fracture energy, $\Gamma_{\rm D}$, in predominantly rate-independent soft materials. Yet, many soft materials exhibit rate-dependence, which gives rise to new physical effects. The simplest rate-dependent material response is viscoelastic, which for highly deformable soft materials is generally nonlinear. Since nonlinear viscoelasticity poses great technical difficulties, mainly because the viscoelastic relaxation times themselves depend on the stress/strain history, we focus here on linear viscoelasticity with the aim of highlighting physical effects that have not been discussed above. The essence of viscoelasticity is the existence of intrinsic material time scales, which were excluded from our discussion so far. In the simplest case, which we consider here, a linear viscoelastic material is characterized by a single relaxation time scale $\tau$. The inverse time scale $1/\tau$ (rate) separates two limiting elastic (non-dissipative) behaviors, whereas dissipation occurs at intermediate rates. In particular, for strain-rates much smaller than $1/\tau$, the elastic response is characterized by a Young's modulus $E_\infty$ (long-time response), while for strain-rates much larger than $1/\tau$, the elastic response is characterized by a Young's modulus $E_0$ (short-time/instantaneous response). Typically, $E_0/E_\infty\!\gg\!1$, i.e.~the short-time elastic response is much stiffer than the long-time one.

To understand the behavior of cracks in materials that are characterized by such a response --- and in particular the crack tip behavior, associated length scales and fracture energy ---, consider a crack steadily propagating at a non-inertial speed $v\!\ll\!c_s$ under static tensile loading, see {\bf Figure~\ref{fig:bulk_dissipation}a}. Assume then that viscoelastic flow is confined to a small region near the tip, and that material far away from this region is fully relaxed and characterized by the softer modulus $E_\infty$, as is depicted on the left part of {\bf Figure~\ref{fig:bulk_dissipation}b}. Under steady-state crack propagation conditions, the stress/strain fields are dragged with the crack tip at a speed $v$, giving rise to a strain-rate distribution that increases as the tip is approached. In linear viscoelasticity, the strain/stress distribution near the crack tip is exactly the same as in LEFM, i.e.~it follows the universal singularity $\sim\!K/\sqrt{r}$~\cite{persson2005crack, Knauss1973, Schapery1975, deGennes1996}. Consequently, the strain-rate follows a singular behavior $\sim\!vK/r^{3/2}$, which implies that the immediate vicinity of the crack (near the yellow region on the left part of {\bf Figure~\ref{fig:bulk_dissipation}a}) is characterized by the short-time/instantaneous modulus $E_0$. Hence, the near tip energy balance condition is $G\!\sim\!K^2/E_0\!\sim\!\Gamma_0$, where $\Gamma_0$ is the intrinsic fracture energy (characterizing the crack tip dissipation in the yellow region). The relation $G\!\sim\!K^2/E_0\!\sim\!\Gamma_0$ is, however, independent of the crack speed $v$, in sharp contrast with the general expectation for a rate-dependent material and with experimental observations~\cite{deGennes1996}. This apparent paradox/contradiction has been identified and discussed in the literature for quite some time, see for example~\cite{Rice1978,Knauss2015}.

The crux of the paradox is related to the main theme of this review, i.e.~to the absence of intrinsic length scales in conventional theories of fracture. In more mathematical terms, in a scale-free theory such as linear viscoelasticity, there exists no length scale that can make $v\tau$ nondimensional, hence there is no way to properly account for the competition between the relaxation time $\tau$ and the crack propagation speed $v$. Indeed, it has been shown~\cite{Knauss1973, deGennes1996} that the introduction of a length scale associated with the intrinsic fracture energy $\Gamma_0$, conventionally termed the cohesive zone regularization length and denoted as $l_c$, resolves the paradox. In the limit of small propagation speeds, $v\tau/l_c\!\ll\!1$, the material responds as an elastic (rate-independent) material with a modulus $E_\infty$ and fracture energy $\Gamma\!\approx\!\Gamma_0$. This implies that $\Gamma_{\rm D}$ is negligibly small --- i.e.~that viscoelastic bulk dissipation is negligible ---, that the crack tip energy balance reads $G\!\sim\!K^2/E_\infty\!\sim\!\Gamma_0$ and that $l_c$ can be identified with the dissipation length $\xi$.

The situation is qualitatively and quantitatively different in the opposite limit of high propagation speeds (yet still non-inertial ones), $v\tau/l_c\!\gg\!1$. In this limit, viscoelastic dissipation is maximal and $\Gamma_{\rm D}\=\Gamma-\Gamma_0\=K^2/E_\infty-K^2/E_0\=(E_0 \Gamma_0/E_\infty)-1\!\approx\!E_0\Gamma_0/E_\infty\!\gg\!\Gamma_0$~\cite{deGennes1996, Hui1992}. That is, the fracture energy is dominated by viscoelastic dissipation and the dissipation zone is much larger than $l_c$ (by the same enhancement factor $\Gamma/\Gamma_0\!\gg\!1$). Consequently, the fracture energy does depend on the propagation speed, $\Gamma(v)$, varying monotonically between the two limits
\begin{equation}
\Gamma(v\!\ll\!l_c/\tau) \to \Gamma_0\ll\Gamma_{\rm D} \qquad\qquad\hbox{and}\qquad\qquad \Gamma(v\!\gg\!l_c/\tau)\approx \Gamma_{\rm D} \to E_0\Gamma_0/E_\infty \gg \Gamma_0 \ ,
\end{equation}
which is qualitatively consistent with the general expectation and resolves the paradox.

When the crack propagation speed $v$ is not very small, the crack tip region features a ``trumpet'' structure~\cite{deGennes1996, Saulnier2004}, as illustrated on the left part of {\bf Figure~\ref{fig:bulk_dissipation}b}. In the immediate tip region, i.e.~at a distance $r$ that satisfies $l_c\!<\!r\!<\!v\tau$, the material behaves as a relatively hard elastic solid characterized by an elastic modulus $E_0$. Next to it, for $r$ in the range $v\tau\!<\!r\!<\!E_0 v \tau/E_\infty$, the material behaves as a viscous liquid and dissipates energy. Finally, for $r\!>\!E_0 v \tau/E_\infty$ the material behaves as a relatively soft elastic solid characterized by an elastic modulus $E_\infty$. This ``viscoelastic trumpet'', which reflects the spatial variation of the local strain-rate, implies that material points located at different distances away from the crack line experience different loading-unloading histories as the crack propagates. This is illustrated on the right part of {\bf Figure~\ref{fig:bulk_dissipation}b}, which reveals location-dependent hysteresis loops that vary even more pronouncedly than in the rate-independent case of {\bf Figure~\ref{fig:bulk_dissipation}a}. It is important to note that the physical picture discussed above formally pertains to infinite systems, and that the intervention of finite geometrical length scales --- such at the system height $H$ --- may enrich the picture further (e.g.~giving rise to a non-monotonic $\Gamma(v)$~\cite{Saulnier2004, Xu1992}).

The discussion in this subsection made several simplifying assumptions about the viscoelastic material response, most notably that it is linear and characterized by a single time scale. While allowing us to various basic concepts and physical effects, these assumptions --- mainly that of linearity --- may fall short of quantitatively accounting for the fracture of rate-dependent soft materials, such as polyampholyte gels~\cite{Sun2013}, which can support strains above 1000\%. Although many nonlinear viscoelastic models have been proposed in the literature~\cite{Wineman2009, Bergstrom1998, Vernerey2017, Mao2017}, only few of them have been applied to the fracture highly deformable soft materials~\cite{Guo2018, Liu2019} due to the technical difficulties involved and the uncertainties in the material parameters. Nonlinear viscoelasticity, which combines geometric nonlinearity with rate-dependence, also implies that the extensively discussed length scales $\xi$ and $\ell$ may no longer be well-defined. These challenging situations lie at the forefront of current research on the fracture of highly deformable soft materials.

\section{Conclusions and open challenges}
\label{sec:conclusions_challenges}

In this review, we highlighted two length scales $\ell$ and $\xi$ that underlie the fracture of highly deformable soft materials. The nonlinear elastic length scale $\ell$ represents the distance from a crack tip below which nonlinear elastic fields, associated with large material deformation, are dominant. The dissipation length scale $\xi$ represents the size of a region near the crack tip where singular fields are no longer dominant, in which material failure is governed by microstructural details and local statistical processes.

Our understanding of the nature of nonlinear elastic crack tip fields is mostly restricted to 2D. Very little is known about the nature of the asymptotic nonlinear elastic fields in 3D, for example, when the crack front (instead of the crack tip in 2D) is curved. Moreover, we focused in this review on tensile (mode-I) fracture as most experiments are conducted under such symmetry conditions. Yet, while nonlinear elastic crack tip field solutions under mixed mode conditions (i.e.~situations involving both tensile and shear loading) are available~\cite{Long2015}, relevant experimental work falls far behind. Indeed, much more work is needed in order to understand the effect of mixed mode loading on crack initiation and growth in highly deformable soft materials.

The length scales $\ell$ and $\xi$ are no longer strictly well-defined in rate-dependent, highly deformable soft materials such as self-healing hydrogels~\cite{Sun2013}. In these materials, the work of extension $W_*$ depends on the loading rate and Young's modulus $E$ in Equation~\ref{eq:ell} may vary by three orders of magnitude with varying strain-rate. How rate-dependent material behavior affects fracture poses one of the most important and challenging open problems in the field. Most currently available theories are based on linear viscoelasticity, which assumes small deformation and that every material point exhibits the same relaxation behavior, independently of the stress/strain history. Nonlinear viscoelasticity, like plasticity in metals, may introduce new length scales into the fracture problem (e.g.~the size of the plastic zone), which may potentially shed light on unresolved problems in linear viscoelastic fracture theory, see for example~\cite{Knauss2015, Gent1996}.

Most (but not all) of the discussion above has been couched within a continuum framework, invoking coarse-grained quantities such as $\Gamma$ and $W_*$. Yet, the continuum approach is expected to break down near the crack tip, where the material structure/architecture plays important roles. Indeed, understanding how the structure/architecture of the underlying polymeric network controls fracture-associated energy dissipation is extremely important for understanding soft materials failure and for material design. As discussed extensively above, highly deformable soft materials can feature a large dissipation length $\xi$, where damage accumulates through chain scission and related molecular processes (e.g.~in self-healing gels, where some chains in the network are connected by physical cross-links, chains can also heal). The distribution of discrete breaking events is controlled by local load transfer, which in turn is governed by the network structure and chain dynamics.

Consequently, mean-field/coarse-grained approaches should be supplemented by statistical approaches to failure, in order to relate continuum quantities such as $\Gamma$ and $W_*$ to the underlying network structure. The development of such statistical approaches can be guided by novel experimental techniques, for example fluorescent mechanochemistry that can probe chain scission and local damage near the crack tip~\cite{Ducrot2014, Millereau2018}. Understanding and predicting the strong coupling between continuum descriptions and local/statistical failure mechanisms, as encapsulated in the relation between $\Gamma$ and $W_*$, is a grand challenge in developing a theory of the fracture of highly deformable soft materials.

\section*{DISCLOSURE STATEMENT}
The authors are not aware of any affiliations, memberships, funding, or financial holdings that might be perceived as affecting the objectivity of this review.

\section*{ACKNOWLEDGMENTS}
E.B.~acknowledges support from the Ben May Center for Chemical Theory and Computation and the Harold Perlman Family. R.L.~acknowledges support from a CAREER award from the National Science Foundation (CMMI-1752449). C.Y.H.~acknowledges support from the National Science Foundation (NSF), USA MoMS program under grant number 1903308. J.P.G.~acknowledges support from the Japan Society for the Promotion of Science (JSPS) KAKENHI (grant no.~JP17H06144) and the Institute for Chemical Reaction Design and Discovery (ICReDD) by World Premier International Research Initiative (WPI), MEXT, Japan. We thank Xiaohao Sun and Yuan Qi for their assistance with the preparation of the figures.


\end{document}